\documentclass{article}

\usepackage{PRIMEarxiv}
\usepackage{amsmath}
\usepackage[utf8]{inputenc} 
\usepackage[T1]{fontenc}    
\usepackage{hyperref}       
\usepackage{url}            
\usepackage{booktabs}       
\usepackage{amsfonts}       
\usepackage{nicefrac}       
\usepackage{microtype}      
\usepackage{lipsum}
\usepackage{fancyhdr}       
\usepackage{graphicx}       
\graphicspath{{media/}}     

\usepackage{graphicx}
\usepackage{caption}

\title{A Shortcut to Structure in AlphaFold 3

}

\author{
Jonathan Feldman\textsuperscript{1,2,3},
Jeffrey Skolnick\textsuperscript{2,3, 4*} \\
\textsuperscript{1}College of Computing, Georgia Institute of Technology, Atlanta, Georgia, USA \\
\textsuperscript{2}Center for the Study of Systems Biology, Georgia Institute of Technology, Atlanta, Georgia, USA \\
\textsuperscript{3}School of Biological Sciences, Georgia Institute of Technology, Atlanta, Georgia, USA \\
\textsuperscript{4}Parker H. Petit Center for AI-Driven Health Innovation, Atlanta, Georgia, USA \\
\textsuperscript{*}Corresponding author: \texttt{skolnick@gatech.edu}
}

\begin{document}
\maketitle

\begin{abstract}

AlphaFold 3 predicts protein structures with remarkable accuracy, yet how structural information emerges within the model remains poorly understood. Here, through causal interventions on internal representations and direct probing of every Pairformer block, we trace the formation of global protein geometry and identify the multiple sequence alignment (MSA) as a structural shortcut to the fold. Removing the MSA largely preserves local secondary structure while disrupting the long-range relationships that define global topology. Restoring the MSA-enriched pair representation at only forty residues recovers most of this lost organization, including at pairs never directly modified. This contribution depends on the detailed direction of the MSA module’s output rather than its magnitude. The Pairformer rapidly converts this signal into global geometry: the final fold becomes recoverable by approximately block 9 of 48 for a majority of proteins, roughly twenty-seven blocks before the model’s decoder can render it, whereas without the MSA it remains inaccessible for most proteins throughout the pass. Which homologs are supplied shapes this trajectory more strongly than which query is supplied; it persists for a designed query that never evolved but collapses for a shuffled sequence. Most importantly, an alignment built for a different protein that shares the fold, supplied only at the structurally corresponding columns, raises the median TM-score against experiment from 0.44 to 0.72, while the same alignment shifted a few residues along the chain performs worse than supplying no alignment at all. What AlphaFold 3 reads from an alignment is therefore a description of the fold itself, transferable between proteins that share one, rather than the query's own evolutionary history. This explains both its accuracy and the limits of what it has solved.

\end{abstract}


\section{Introduction}

\begin{figure}[t]
\centering
\includegraphics[width=\textwidth]{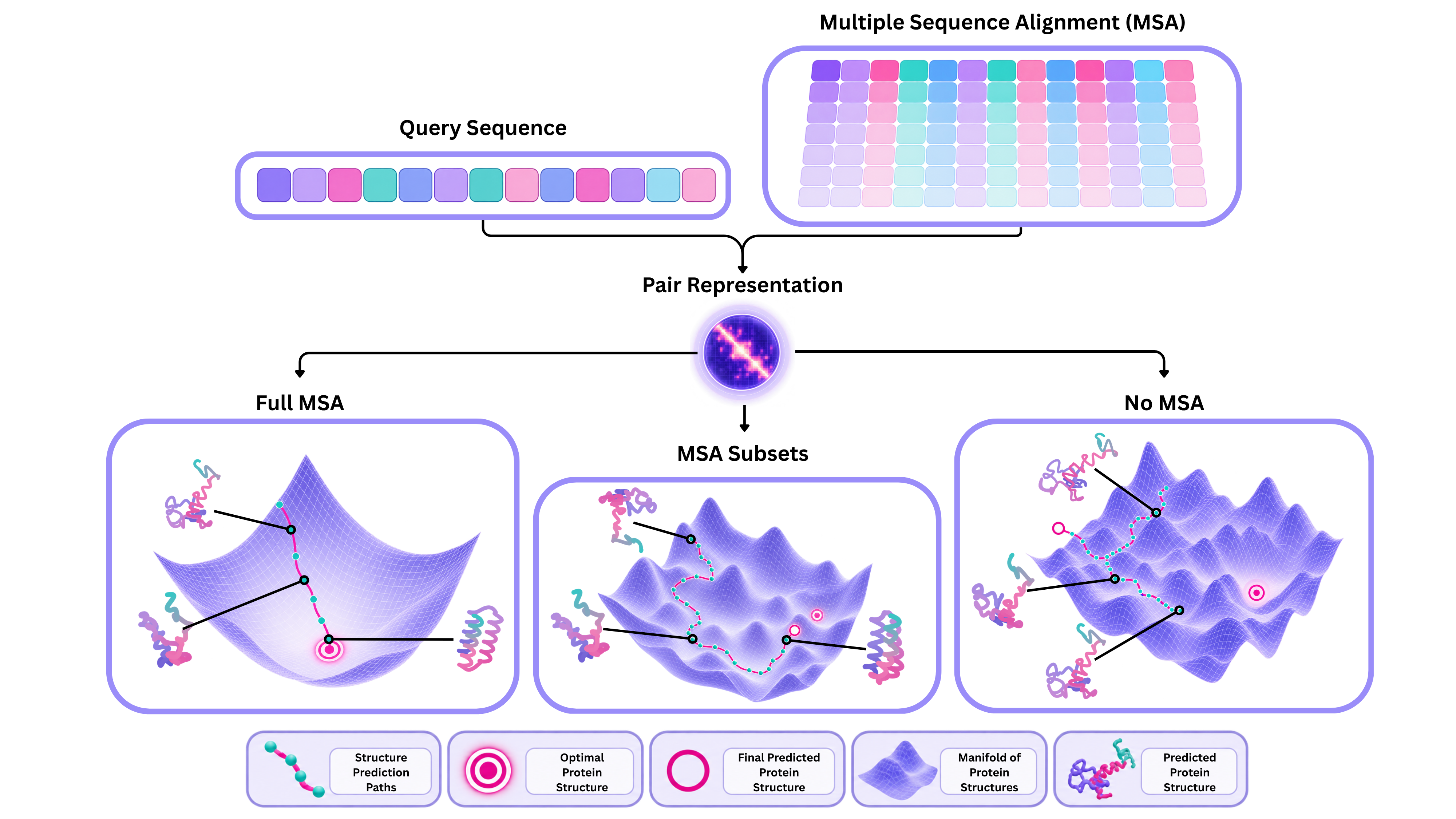}
\caption{\textbf{Schematic of the alignment as a structural shortcut to the fold.}
As shown, the query sequence and its multiple sequence alignment (MSA), which provides evolutionary information from homologous proteins, are combined into the pair representation, which the Pairformer then transforms over 48 blocks. In this schematic, the difficulty of extracting the accurate fold is conveyed by the ruggedness of a terrain that the model traverses, and each point along a path marks one Pairformer layer. With a full alignment, the terrain is a smooth funnel and the global fold is already decodable from the representation after ten of the forty-eight blocks. With a subset of five distant homologs, the terrain is more rugged, and the same fold, though less accurate, is reached late in the pass by a far longer route. Without an MSA, the terrain offers no clear descent, and for many proteins the correct global fold is never reached at all. The structures shown along each path are stand-ins for the kind of prediction one would expect at that stage, in relative quality against the experimental structure, rather than readouts from any individual protein. What flattens the terrain is not the alignment's evolutionary relationship to the query but the structure it encodes. Evolutionary information therefore does not merely improve the prediction---the structure latent within it is what makes the fold accessible, and how much of it the model receives sets how quickly the structure appears. This schematic is informative by analogy: there is no indication that AlphaFold 3 optimizes over a loss landscape, and the terrain is a proxy for the difficulty of the underlying problem.}
\label{fig:schematic}
\end{figure}

The release of AlphaFold 2 in 2021 transformed structural biology. For the first time, it was possible to predict protein structures with near-experimental accuracy for many targets \cite{jumper_highly_2021,abramson_accurate_2024,tunyasuvunakool_highly_2021, yang_alphafold2_2023, 10.1093/bioinformatics/10.6.587, BETANCOURT2004635}. Its release also catalyzed a wave of new research combining biological and AI techniques \cite{thadani_learning_2023, feldman_biophysically_2025, cho_protein_2025, stark_boltzgen_2025, pacesa_one-shot_2025, hekkelman_alphafill_2023, Kalakoti2025-rf}. In 2024, AlphaFold 3 extended these capabilities to nucleic acids, small molecules, and biomolecular assemblies, expanding an already powerful tool into one suitable for protein design, drug discovery, and biosecurity \cite{roy_alphafold3_2024, yang_alphafold2_2023, skolnick_alphafold_2021, feldman_af3complex_2025, gao_af2complex_2022, Zhao2025.05.07.652715, feldman_resilient_2025, dianzhuo_wang_without_2025, Feldman2026-xy}. Yet, in spite of these advances, two critical questions have persisted: how is AlphaFold able to transform an amino-acid sequence into an accurate three-dimensional structure, and to what extent does its prediction process reflect the biophysics underlying protein folding?

In our previous work, we set out to answer these questions \cite{alphainterp}. In that work, we examined the internal representations of AlphaFold 3 in a novel mechanistic interpretability study, demonstrating that removing the multiple sequence alignment (MSA), which provides evolutionary information from homologous proteins \cite{roy_alphafold3_2024,rustamov_msa_2024, biom14121531}, substantially deteriorates AlphaFold 3's internal representations, confidence, and structural predictions, while reintroducing as few as one to five sufficiently diverse homologs can recover much of the benefit of a full MSA \cite{alphainterp, feldman_adversarial_2026}. However, our findings left several important questions unanswered. The most important among them are: Which structural relationships depend most strongly on evolutionary information, and is this related to the evolutionary history of the protein or the particular fold it adopts? How much of the MSA-enriched representation is required to recover the global fold? And how does the Pairformer transform this information into structural geometry?

Addressing these questions requires looking beyond the final structures AlphaFold 3 outputs and examining the internal representations and computations that produce them. We proceed in five steps. First, as an observation, we identify which features of a predicted structure depend on evolutionary information and which survive its removal. Second, through causal intervention, we show that restoring a sparse portion of the MSA-enriched representation recovers much of the lost global structure, including relationships never directly restored. Third, to identify the mechanism, we follow this representation through the Pairformer and show that, although the representation entering the Pairformer yields no global fold under linear decoding, the fold becomes accessible within the first ten of its 48 blocks for over half of proteins---whereas without the MSA, it does not emerge for most proteins within the first pass. Fourth, to establish the source of this information, we independently vary the MSA and the query sequence, and examine sequence-divergent proteins that share a fold, revealing that the MSA sets when and along which route the fold emerges, while the query determines whether it emerges at all. Fifth, and most directly, we transplant the structurally aligned columns of one protein's alignment into a sequentially dissimilar protein that shares its fold, and find that this alone recovers much of the structure, while the same alignment displaced by a few residues performs worse than supplying none. Together, these steps reveal a structural shortcut at the center of this study: what AlphaFold 3 reads from an alignment is a description of the fold, supplied in the correct correspondence, rather than the query's evolutionary history.

\section{Results}

\subsection{MSA information preferentially supports nonlocal structural organization}

We began with the two cohorts of 200 monomeric proteins assembled in our previous work \cite{alphainterp}: novel proteins without a match above 40\% sequence identity to any sequence deposited before AlphaFold 3's September 30, 2021 training cutoff, and similar proteins exceeding that threshold. All were themselves deposited after the cutoff (see Methods).

We generated AlphaFold 3 predictions with and without an MSA and compared them with the corresponding experimental structures from the PDB to quantify the loss in accuracy. Separately, we compared each no-MSA prediction against its corresponding full-MSA prediction, allowing us to identify which features of the model's preferred structural solution were retained or lost following the removal of what at first glance appears to be evolutionary information. As expected from our previous findings \cite{alphainterp}, the presence of evolutionary (or perhaps fold) information as encoded in the MSA contributed substantially to predictive accuracy in both cohorts, with the median TM-score \cite{zhang_tm-align_2005} against experiment declining from 0.96 to 0.52 in novel proteins and from 0.97 to 0.46 in similar proteins following MSA removal (Supplementary Figure~S1A). These results establish the magnitude of the structural deterioration caused by removing the MSA, but do not explain what structural information is lost or why this deterioration occurs.

To identify which features were lost, we examined their retention following MSA removal and found a pronounced difference between local secondary structure and nonlocal relationships (Supplementary Figure~S1B). In the novel cohort, 84\% of helical residues and 71\% of strand-like residues present in the full-MSA predictions were retained without the MSA. In contrast, exact $\beta$-bridge \cite{Bhat2020-sn} retention declined to only 42\%, and contact retention declined with sequence separation, reaching 45\%, 40\%, and 28\% at short, medium, and long range. The similar cohort behaved relatively uniformly throughout (Supplementary Section~S1). Note that secondary-structure retention is a per-residue measure and does not require those elements to occupy the same positions within the three-dimensional fold, so a helix may remain recognizable even where it is incorrectly placed in the global structure.

Thus, although much of the local secondary structure predicted by AlphaFold 3 can survive MSA removal, the much greater loss of exact $\beta$-bridge partnerships and long-range contacts is indicative of the fact that the MSA is of immense importance when establishing the nonlocal relationships required to organize protein structural elements into a geometrically consistent three-dimensional fold.

\begin{figure}[p]
\centering
\includegraphics[width=\textwidth,height=0.72\textheight,keepaspectratio]{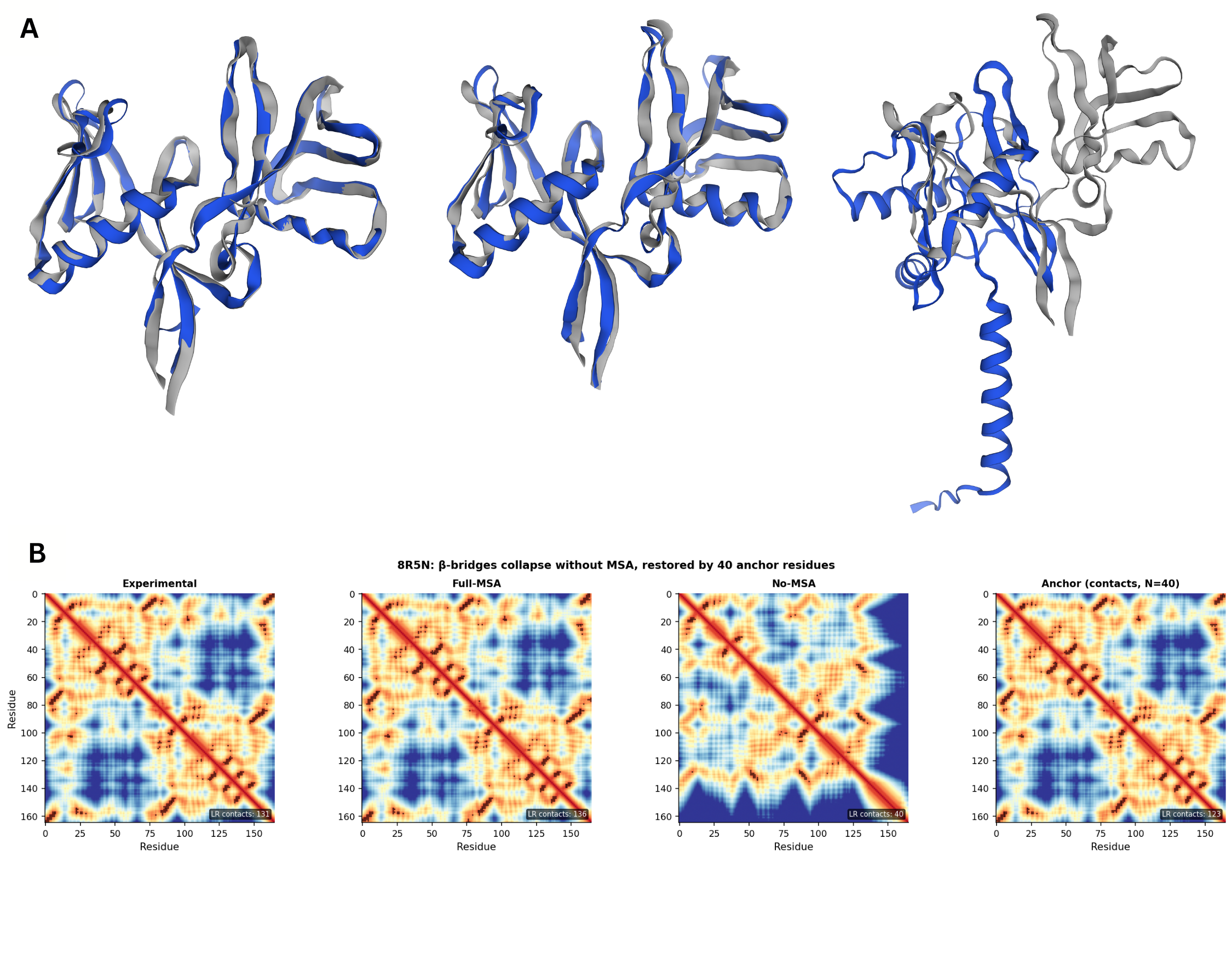}
\caption{\textbf{Sparse restoration of the MSA recovers long-range structural organization in 8R5N.}
\textbf{(A)} Structural overlays against experiment (gray) for the full-MSA prediction, the 40-residue anchor intervention, and the no-MSA prediction (blue). Corresponding TM-scores against experiment are 0.984, 0.961, and 0.261.
\textbf{(B)} C$\alpha$ distance matrices for the experimental structure, full-MSA prediction, no-MSA prediction, and anchor intervention. Long-range contact counts are 131, 136, 40, and 123, respectively. Exact $\beta$-bridge partnerships are lost without the MSA and recovered following anchor restoration.}
\label{fig:structural_example}
\end{figure}

This distinction is illustrated by the representative novel protein 8R5N. Its full-MSA prediction closely reproduced the experimental structure (TM-score = 0.984), whereas removing the MSA reduced the TM-score to 0.261 (Figure~\ref{fig:structural_example}A). Although recognizable local elements remained present, the prediction failed to recover the correct $\beta$-sheet organization, losing every exact $\beta$-bridge partnership present in the full-MSA prediction and substantially disrupting long-range residue relationships (Figure~\ref{fig:structural_example}B). The full-MSA prediction contained 136 long-range contacts, closely matching the 131 present in the experimental structure. The no-MSA prediction contained only 40.

\subsection{Sparse restoration of MSA-enriched information recovers the global fold}

We next asked whether recovering the nonlocal relationships lost following MSA removal required the entire MSA-enriched representation or whether restoring only a portion would suffice. To investigate this, we performed activation patching at checkpoint B, immediately after the MSA module and before the Pairformer. For each protein, we obtained pair representations from both a full-MSA and a no-MSA run, and calculated their difference, denoted $$\Delta_B = P_B^{\mathrm{MSA}} - P_B^{\mathrm{noMSA}}.$$ We selected residues, termed anchors, and restored their corresponding rows and columns from the full-MSA pair representation within the no-MSA state. Computation then proceeded normally through the Pairformer and diffusion module (see Methods). Critically, selecting a single anchor does not restore information associated with that residue alone, but also restores its pairwise representations with every other residue in the protein. Hence, our aforementioned intervention restores part of the broader representational context in which the selected residue exists, rather than simply inserting information about an isolated structural feature or contact.

Structural recovery increased with the number of anchors restored, whether specified as a fixed count or as a proportion of sequence length, and under every selection strategy we tested (Supplementary Section~S2). Remarkably, restoring the pair information associated with only 40 contact-selected residues recovered a median of 71\% of the available experimental TM-score gap between the no-MSA and full-MSA predictions among the 328 proteins for which that gap exceeded 0.1. The improvement extended beyond global structural similarity to the nonlocal relationships most sensitive to MSA removal: in the novel cohort, long-range-contact retention rose from 28\% without the MSA to 73\% with 40 contact-selected anchors, relative to the full-MSA prediction. It also extended beyond the pairs directly restored, reaching those residue pairs whose representations were never explicitly enriched (Supplementary Section~S2).

The success of these sparse interventions led us to ask whether recovery depended on a particular, privileged collection of anchor residues, or whether several entirely different subsets could provide sufficient information to reconstruct the fold. To investigate this, we constructed four mutually disjoint sets of 40 residues for each protein, using both contact-based and conservation-based selection, and independently repeated the restoration intervention for each set (Supplementary Figure~S1D; see Methods). Despite containing different residues, the four contact-selected sets produced comparable TM-scores of 0.80, 0.81, 0.81, and 0.82 against experiment, while the four conserved sets produced 0.78, 0.80, 0.79, and 0.80. Recovery therefore did not depend on one indispensable anchor set. The information the MSA supplies is distributed and partially redundant, and a small fraction of it suffices to rebuild the fold.

\subsection{The MSA module writes a distributed evolutionary representation that influences structural prediction}

Having shown that sparse restoration of the MSA-enriched state can causally rebuild the fold, we next asked what the MSA module actually writes into the pair representation.

Our anchor interventions used the full-MSA minus no-MSA difference at checkpoint B, but this difference may also include contributions inherited from earlier stages. We therefore calculated the module updates within each run and defined their difference as the MSA-specific module contribution, $$M=(P_B^{\mathrm{MSA}}-P_A^{\mathrm{MSA}})-(P_B^{\mathrm{noMSA}}-P_A^{\mathrm{noMSA}}),$$ distinct from the total checkpoint-B difference $\Delta_B$ used above (see Methods). Across the 400 proteins examined, the median projection coefficient of $M$ onto $\Delta_B$ was 0.995, against 0.005 for the difference already present at checkpoint A, and the two contributions were nearly orthogonal. The MSA module writes essentially all of what distinguishes the two conditions.

Two properties of what it writes bear on the results that follow. First, the contribution is organized simply in space, with a median effective rank \cite{roy_effective_2007} of 1.15, and larger updates at positions with greater evolutionary variability and alignment coverage (Supplementary Section~S3). Second, and more consequentially, what matters structurally is not the strength of this contribution but its direction. Disrupting cross-column relationships by shuffling amino acids within alignment columns left the overall orientation of the representation largely intact (cosine similarity 0.938 after shuffling 99\% of eligible columns) while substantially changing the resulting predictions (contact-map Jaccard index 0.740 against the unshuffled prediction). Decomposing the native and shuffled module updates into magnitude and direction and recombining them separates the two: native direction with shuffled magnitude yielded a median TM-score of 0.96 against the full-MSA prediction, identical to the fully native control, whereas native magnitude with shuffled direction yielded only 0.70, barely above the 0.68 of the fully shuffled condition. Per protein, restoring the native direction recovered a median of 1.000 of the available structural difference; by contrast, restoring the native magnitude recovered 0.011 (Supplementary Section~S3).

Representational similarity therefore need not imply functional equivalence: two representations differing by a cosine distance of only 0.06 can yield substantially different structures, because the difference lies in the directions that matter. What the MSA supplies is carried in the detailed pattern of the representation rather than its overall strength.

\subsection{The Pairformer transforms the MSA into structural geometry}

If the MSA module writes the information that rescues structure, the next question is how the Pairformer transforms that signal into a representation of geometry.

We thus followed the pair representation through all 48 blocks of the first Pairformer pass, comparing full-MSA and no-MSA representations at the same block. This allowed us to distinguish the component that remained aligned with the original MSA-specific contribution from changes in representational form, and to examine when structural information became accessible (see Methods).

\begin{figure}[p]
\centering
\includegraphics[width=\textwidth,height=0.72\textheight,keepaspectratio]{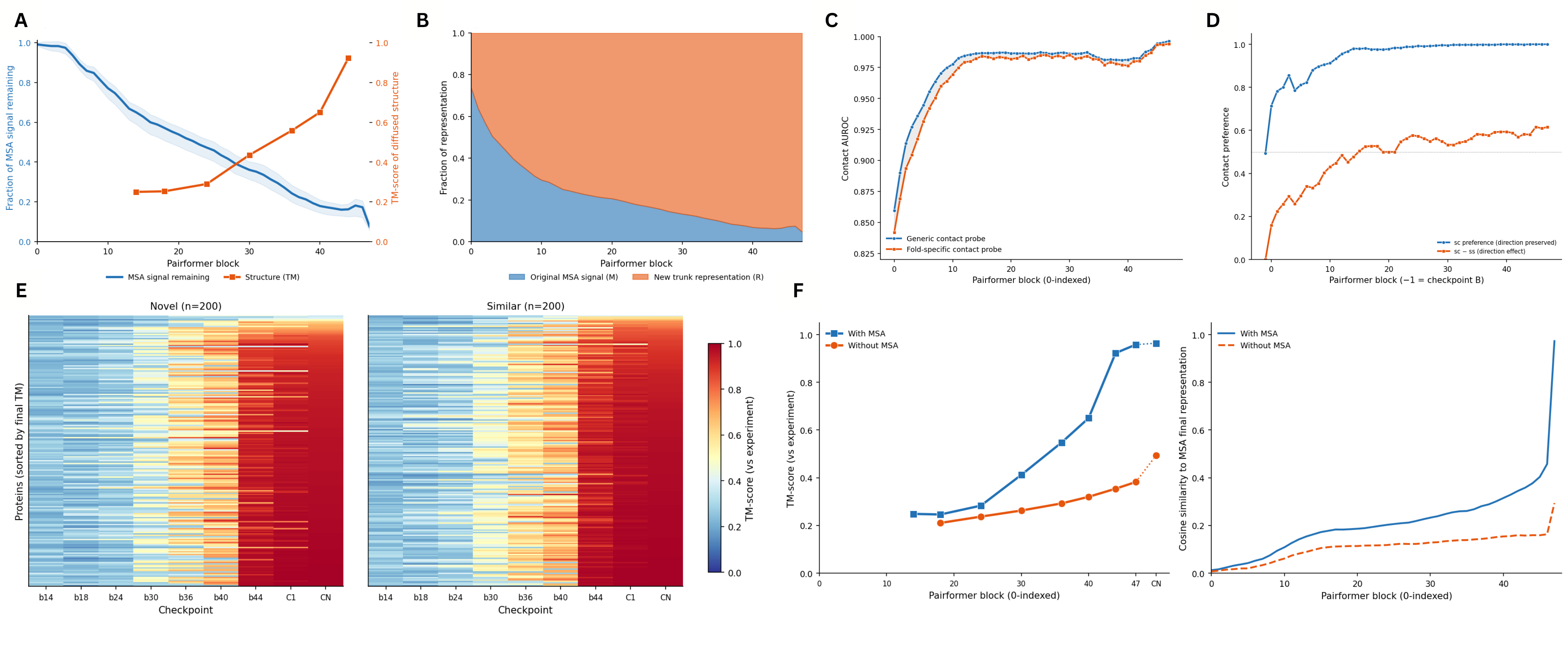}
\caption{\textbf{The Pairformer transforms the MSA into structural geometry.}
\textbf{(A)} Projection of the evolving MSA-dependent representation onto the original MSA-module contribution (blue), alongside intermediate experimental TM-score (orange).
\textbf{(B)} Relative contributions of the original MSA-aligned component and the orthogonal residual.
\textbf{(C)} Generic and fold-specific contact-probe AUROC \cite{Fawcett2006-wf} across Pairformer blocks.
\textbf{(D)} Native-contact preference with native direction preserved (blue) and its difference from the fully shuffled condition (orange). Their respective baselines are 0.5 and 0.
\textbf{(E)} Per-protein experimental TM-scores across intermediate checkpoints in the novel and similar cohorts ($n=200$ each).
\textbf{(F)} Left: median TM-score against experiment for structures diffused independently from the saved representations at each checkpoint, with and without the MSA (see Supplementary Information Section S4). Right: cosine similarity between the pair representation at each block and the final representation of the same run. The late structural improvement is substantially greater with the MSA, while the final Pairformer block produces a pronounced representational change.}
\label{fig:emergence}
\end{figure}

We first decomposed the difference 
$$P_{\mathrm{MSA}}^{(l)}-P_{\mathrm{noMSA}}^{(l)}$$\\ at each block into its projection onto the original MSA-specific contribution, denoted $M$, and an orthogonal residual, denoted $R$. This allowed us to distinguish the component that remained aligned with the initial contribution from the remainder of the representation (Figure~\ref{fig:emergence}B; see Methods). As established in our previous analysis, the MSA-aligned component declined progressively as we passed through the Pairformer, retaining approximately 54\% of its initial value by block 20 and only 8\% by block 47. In contrast, the orthogonal component continued to grow and became dominant around block 25.

We also observed increasing contact enrichment within this residual representation between blocks 20 and 40 (see Methods). Let us briefly note that the decline of the original MSA-aligned component does not necessarily mean that the MSA-encoded information has disappeared, nor does it mean that the orthogonal residual represents information independent of the MSA. Rather, the Pairformer may be transforming the original evolutionary contribution into a different representational form as it processes and enriches the pair representation. These observations demonstrate that substantial representational reorganization occurs well before an accurate structure can be generated.

As this representation reorganizes (Figure~\ref{fig:emergence}C), an important question is when contact information becomes decodable. In other words, at what point do the intermediate pair representations contain information about which residues will be in proximity to one another in the final predicted structure? 

To examine this, we trained contact probes on separate proteins and evaluated their ability to identify contacts present in the final full-MSA prediction from intermediate Pairformer representations (see Methods). We evaluated both overall contact-versus-noncontact discrimination and discrimination between contacts unique to the native and shuffled predictions. For the generic probe, contact AUROC was already approximately 0.86 after block 0 and increased rapidly to approximately 0.99 by block 9. The fold-specific probe initially performed somewhat worse, but its performance converged with that of the generic probe by approximately block 15. Thus, contact information was already substantially decodable during the earliest Pairformer computations, and its accessibility improved considerably over the subsequent blocks.

However, high contact AUROC does not necessarily establish that the corresponding residues have been placed at the correct distances in a generated structure, nor does it establish that AlphaFold 3 itself uses this information in the same manner as our probes. Contacts are only one part of a protein structure---they do not fully define the geometric relationships required to construct a fold. These results therefore establish when contact information becomes accessible, not when the complete three-dimensional structure has been resolved. 

Next, we returned to the magnitude--direction interventions from the preceding section to examine when the structural consequences of those perturbations became apparent within the pair representation. We used the same four combinations of native and shuffled MSA-module magnitude and direction and applied our contact probe to the representation generated under each condition at every Pairformer block. 

For each protein, we identified two sets of contacts: native-only contacts, which were present in the native-MSA prediction but absent from the shuffled-MSA prediction, and shuffled-only contacts, which were present in the shuffled prediction but absent from the native prediction. We then measured how often the probe assigned a higher score to a native-only contact than to a shuffled-only contact. We matched the compared residue pairs by sequence separation to account for differences in contact-prediction difficulty (see Methods). We refer to this separation-matched ranking statistic as the control preference. A value of 0.5 indicates chance-level discrimination, while a value of 1.0 indicates that native-only contacts are always ranked higher.

At checkpoint B, the native representation exhibited a preference of only 0.494 (Figure~\ref{fig:emergence}D), approximately equal to chance, despite the presence of generally decodable contact information. Following the first Pairformer block (block 0), however, preference increased to 0.714, reached 0.782 after block 1, and approached 1.0 by block 15. The four magnitude--direction conditions also allowed us to examine the contribution of representational direction to this preference. The difference in control preference between the shuffled-magnitude/native-direction and fully shuffled conditions ($\mathrm{sc}-\mathrm{ss}$) was 0.159 after block~0 and continued to rise through subsequent blocks (see Methods). 

Hence, even though the native and shuffled MSA-module contributions already contained differences capable of influencing the final structural predictions, the contacts that distinguished those predictions were not preferentially identified by our probe at checkpoint B. This distinction became apparent during the early Pairformer layers and increased substantially thereafter. 

\subsection{Intermediate diffusion reveals an overcompact proto-fold}

The representational analyses above establish that the MSA-derived state is rapidly reorganized, but not when AlphaFold’s own structural decoder can turn that state into coordinates.

To investigate this, we independently generated structures from saved representations at seven Pairformer blocks (14, 18, 24, 30, 36, 40, and 44) and two additional checkpoints ($C_1$ and $C_N$) by passing them through the diffusion module, and compared each against the final full-MSA prediction, $C_N$ (see Methods).

The earliest structural predictions were overly compact and poorly organized. At block~14, the radius of gyration was only 48\% of that of the final prediction, helix and exact $\beta$-bridge recovery were effectively zero, strand recovery was 2\%, and global structural agreement was 6\%. We call this initial state, as generated by the diffusion module, a proto-fold. Through the remainder of the first pass the readouts gradually expanded and recovered recognizable features, but in a strict order: helices first, then strands, and exact $\beta$-bridge partnerships last, with the global architecture resolving only near block~44 (Supplementary Section~S4). This progression parallels our MSA-removal results, where recognizable local secondary structure can be retained or recovered without the precise nonlocal relationships required for the final fold. The representative protein 8R5N illustrates the same progression (Supplementary Figure~S3).

\subsection{Probing the pair representation reveals an early, MSA-dependent global fold}

Because the diffusion module was trained on completed representations, however, its failure to render an early structure does not establish that the fold is absent from the representation itself.

To distinguish between these possibilities, we examined what the representation itself contains. Because contacts are already linearly decodable from the pair representation early in the Pairformer (Figure~\ref{fig:emergence}C), we asked whether this decodability extends from individual contacts to the structure as a whole.

To do so, we trained linear probes on the 200 proteins of the similar cohort to predict the final C$\alpha$--C$\alpha$ distances directly from the pair representation at checkpoints A and B and after each of the 48 Pairformer blocks, with one probe per layer and separate probes for representations generated with and without the MSA (see Methods). Because probing is far less computationally expensive than diffusion, we could examine every block of the pass rather than a selected subset. We then applied each probe to the corresponding representations of the 200 held-out novel proteins and reconstructed three-dimensional structures from the decoded distances using multidimensional scaling \cite{Pele2011-ag, Sims2005-fq}. Because the final full-MSA predictions of these proteins closely reproduce their experimental structures (median TM-score 0.96), decoding toward them closely approximates decoding toward experiment.

These reconstructions are interpolations of the global arrangement implied by the decoded distances, so their local geometry is imperfect and individual residues deviate from their final positions even in the best cases. Against a generic-chain floor of 0.22 and a residue-permutation control that drives the explained variance below zero (Supplementary Section~S5), they nonetheless recover residue-specific structural information rather than generic distance statistics.

\begin{figure}[p]
\centering
\includegraphics[width=\textwidth,height=0.85\textheight,keepaspectratio]{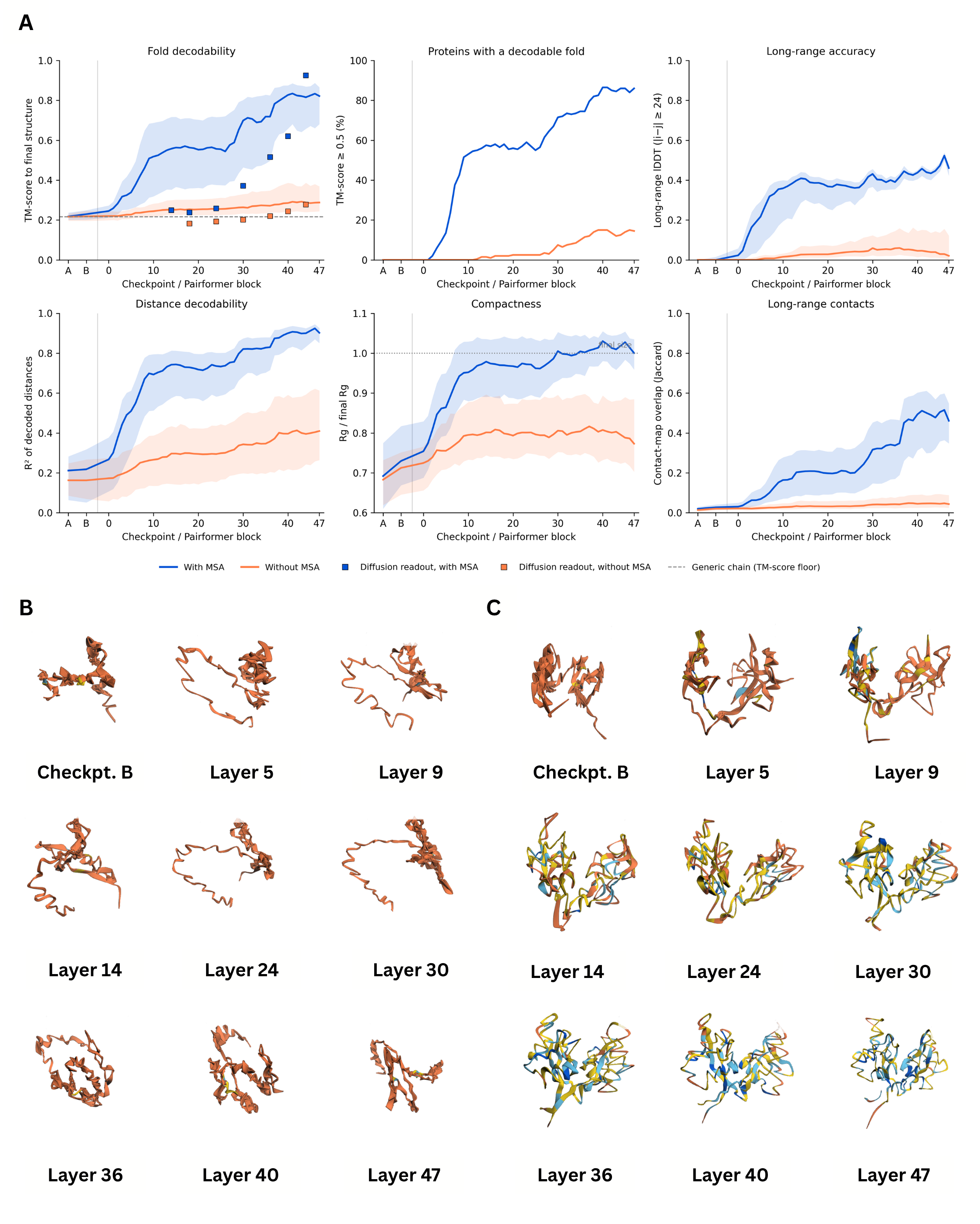}
\caption{\textbf{With the MSA, the global fold becomes decodable from the pair representation within the first ten Pairformer blocks; without it, the fold does not emerge for most proteins.}
\textbf{(A)} Linear probes trained on the similar cohort decode C$\alpha$--C$\alpha$ distances of the final structure from the pair representation at checkpoints A and B and after each of the 48 blocks of the first Pairformer pass, for the 200 novel proteins with the full MSA (blue) or without it (orange). Panels show, from left to right and top to bottom: TM-score of the reconstructed structure against the final structure; the percentage of reconstructions that reach a TM-score of at least 0.5; long-range lDDT ($|i-j| \geq 24$) computed from the decoded distances; the coefficient of determination ($R^2$) of the decoded distances; the radius of gyration of the reconstruction as a ratio of its final value; and the Jaccard overlap between the long-range contact maps of the reconstruction and the final structure. Lines are medians and shaded bands are interquartile ranges. 
\textbf{(B--C)} Structures of 8R5N reconstructed from the decoded distances without the MSA \textbf{(B)} and with it \textbf{(C)}, at checkpoint B and Pairformer blocks 5, 9, 14, 24, 30, 36, 40, and 47. Each residue is colored by the distance of its C$\alpha$ atom from its position in the final structure after superposition: dark blue, $\leq 1$~\AA; light blue, 1--2~\AA; yellow, 2--4~\AA; orange, $>4$~\AA. See Methods for probe training, reconstruction, and display refinement.}
\label{fig:probe}
\end{figure}

The results were striking (Figure~\ref{fig:probe}A). At checkpoint B, the fold was not yet decodable either with or without the MSA, with median TM-scores of 0.23 and 0.22, respectively, practically at the floor. By block~9, however, the median TM-score of structures decoded from the MSA-enriched representation rose to 0.51, and over half of the proteins (52\%) exceeded a TM-score of 0.5. Within the first ten blocks, the representation thus reaches the global fold. Decodability continued to rise in a series of steps, reaching 0.70 by block~30 and 0.82 by block~47. The reconstructions also approached the native size from block~9 onward, with the radius of gyration reaching 95\% of the final value, in contrast to the 48\% observed for the diffused proto-fold at block~14. Although these structures are less refined than the diffused structures, the global fold is present roughly 27 blocks before the diffusion module can produce a structure of comparable accuracy against the same final structure, which it reaches only at block~36.

What is equally striking is that, without the MSA, the global fold largely does not emerge in the first Pairformer forward pass. The median TM-score was 0.22 at checkpoint B, 0.24 at block~10, and only about 0.29 at block~47. No protein reached a TM-score of 0.5 by block~9, and only 14\% did so by block~47, while long-range lDDT remained at or below 0.06 throughout the pass. This is not a failure of the probes. The probe trained on representations without the MSA recovered a TM-score of 0.62 at block~47 when applied to representations generated with the MSA, indicating that it learned features from which the fold can be decoded, but that these features are absent from the representation when the MSA is removed. Without the MSA, the model does eventually approach the correct global fold, but slowly and only approximately. Within the first Pairformer pass the representation almost never yields a decodable fold, ending at 0.29 against the full-MSA prediction, and the diffusion readout at the end of that pass reaches only 0.32. Only through recycling does the prediction approach the fold, reaching 0.45 against the full-MSA prediction and 0.49 against experiment at $C_N$, short of the threshold for a correct fold. With the MSA, the first pass alone reaches 0.98.

Together, these results suggest the shortcut we have been building toward. Before the Pairformer, the contribution of the MSA is not yet apparent as a decodable fold. Within the first few blocks of the 48-block pass, it becomes so, with the Pairformer turning the MSA's contribution into a readable global fold within ten blocks of a single pass, whereas without the MSA, the complete pass rarely does. This is particularly notable given that, in practice, the Pairformer is typically applied repeatedly through recycling, which is seemingly unnecessary for monomers with a robust MSA. The representative protein 8R5N illustrates this contrast (Figure~\ref{fig:probe}B--C). Without the MSA, the reconstructed structures remain far from their final positions at every layer, whereas with the MSA, an increasing fraction of residues approaches the final arrangement from block~14 onward.

Finally, we asked whether any property of the MSA or the structure determined which residues became decodable first. No individual statistic did so strongly: residue burial and alignment coverage showed the strongest associations, covariation and conservation were weak, and all features combined explained little more. Although the earliest-decoded residues were spatially clustered, this clustering was largely explained by their burial, providing no evidence for a distinct folding nucleus (Supplementary Section~S6).

\subsection{Intermediate representations reveal convergent MSA-dependent inference trajectories}

\begin{table}[t]
\centering
\small
\begin{tabular}{lcccccc}
\toprule
\multicolumn{7}{l}{\textbf{(a) Progress toward the fold}} \\
Block & Q (subsets A\&B) & H & MPNN & Shuffled & Q (full MSA) & Q (no MSA) \\
\midrule
14  & 0.322 & 0.329 & 0.327 & 0.200 & 0.688 & 0.245 \\
18  & 0.333 & 0.343 & 0.343 & 0.193 & 0.670 & 0.249 \\
24  & 0.346 & 0.367 & 0.366 & 0.195 & 0.668 & 0.251 \\
30  & 0.450 & 0.461 & 0.499 & 0.191 & 0.756 & 0.262 \\
36  & 0.504 & 0.524 & 0.595 & 0.193 & 0.785 & 0.270 \\
40  & 0.638 & 0.648 & 0.714 & 0.200 & 0.852 & 0.284 \\
44  & 0.671 & 0.683 & 0.731 & 0.210 & 0.849 & 0.277 \\
47  & 0.682 & 0.683 & 0.748 & 0.211 & 0.846 & 0.278 \\
\midrule
\multicolumn{7}{l}{\textbf{(b) Similarity between runs}} \\
Block & QA vs QB & Q vs H & Q vs MPNN & Q vs Shuffled & & \\
\midrule
14  & 0.577 & 0.689 & 0.644 & 0.319 & & \\
18  & 0.599 & 0.694 & 0.637 & 0.306 & & \\
24  & 0.613 & 0.688 & 0.637 & 0.296 & & \\
30  & 0.610 & 0.675 & 0.659 & 0.243 & & \\
36  & 0.663 & 0.686 & 0.673 & 0.230 & & \\
40  & 0.688 & 0.699 & 0.670 & 0.213 & & \\
44  & 0.724 & 0.740 & 0.709 & 0.226 & & \\
47  & 0.750 & 0.748 & 0.721 & 0.244 & & \\
\bottomrule \\
\end{tabular}
\vspace{0.5em}
\caption{\textbf{The MSA sets when and along which route the fold emerges; the query sets whether and how well.}
All values are median TM-scores between structures reconstructed from probe-decoded distances, for 100 novel proteins folded with two evolutionarily independent MSA subsets (A and B) of 5 distant homologs each ($\leq 30\%$ identity to the query and to the other subset). Blocks are numbered 0--47 within the first Pairformer pass.
\textbf{(a)} Reconstructions compared against the final full-MSA prediction of the original query (Q), with the query replaced by a natural homolog (H, $\sim$60\% identity), a ProteinMPNN sequence \cite{Dauparas2022-su} design on Q's backbone (MPNN, $\sim$45\% identity), or Q's residues shuffled, each folded with the same subsets. Full MSA and No MSA are Q's own representations decoded as in Figure~\ref{fig:probe}.
\textbf{(b)} Pairwise TM-scores between reconstructions of the same protein. QA vs QB changes the MSA subset while keeping the query; the remaining columns change the query while keeping the subset (A with A, B with B).
The corresponding diffusion readouts are given in Supplementary Section~S7.}
\label{tab:disjoint}
\end{table}

Having established when the global fold becomes accessible, we next asked which input---the MSA or the query---controls the trajectory by which it emerges.

In the analysis that follows, we varied two things. The first was the MSA, for which we used the two subsets A and B, each of five sequences. The two subsets were constructed to be evolutionarily divergent: every sequence in one subset shares at most 30\% identity with every sequence in the other and with the query, while sequences within a subset share at most 50\% identity with one another. The second was the query, which we either kept unchanged (Q), replaced with a homolog drawn from the protein's own MSA (H), replaced with a structurally equivalent but sequentially divergent sequence designed on the query's own backbone with ProteinMPNN \cite{Dauparas2022-su}, or replaced with a random shuffle of the query's own residues. Every variant was folded with the same two subsets, and the resulting trajectories were examined over the same blocks (Table~\ref{tab:disjoint}).

The diffusion readouts are of limited use here: before about block 40 both subsets produce compact proto-fold blobs of similar shape and size, so their apparent agreement is inflated (Supplementary Section~S7). The probe reconstructions are not subject to this limitation, and they reveal that, with a five-sequence subset, the query hardly matters for the speed of emergence. Q, H, and MPNN move in lockstep, remaining within approximately 0.02 of one another through block 24 and reaching comparable accuracy by the end of the pass, while the shuffled query remains at the floor throughout, as one would expect of a sequence that encodes no structure at all.

What the MSA sets is the speed at which the final structure emerges. With the full MSA, the median TM-score of the reconstructions reaches 0.69 by block 14, whereas with a five-sequence subset it is only 0.32 at the same block and approaches comparable values only near block 40. Without the MSA, it rarely does. Five distant sequences are therefore sufficient to arrive at the global fold, but at a substantially slower rate.

The similarity between runs is a truly key result. Changing the MSA while keeping the query separates the trajectories more than changing the query while keeping the MSA: at block 14, two subsets of the same query agree at a TM-score of 0.58, whereas the same subset decoded from two different queries agrees at 0.69 (Wilcoxon signed-rank $p = 6\times10^{-11}$), and this gap persists to about block 36 before closing by the end of the pass. The trajectory of emergence is thus dictated more by which homologs are supplied than by which query, so long as the query encodes the same general structure in a manner that is physically sensible.

Taken together, these analyses indicate that the MSA sets when the fold emerges and what route it takes in doing so, while the query affects whether it emerges at all and affects how accurate the final structure is. The ProteinMPNN designs are instructive in this regard: although they share their MSA with the original query, they yield the most accurate reconstructions of any perturbed condition we examined (0.75 against 0.68 at the end of the pass). We note that these sequences were designed on the query's own full-MSA backbone, which is also the reference against which they are scored; we would speculate that, having been designed for that backbone directly, they are free of the biological constraints that make native sequences an imperfect encoding of their own structures. And without the MSA, practically nothing emerges at all, once again solidifying its importance and casting light on how powerful it is in determining the fate of a protein structure inside AlphaFold 3.

\subsection{Sequence-divergent homologs converge on a shared fold dictated by the MSA}

\begin{table}[t]
\centering
\small
\begin{tabular}{lcccccc}
\toprule
 & \multicolumn{3}{c}{Full MSA} & \multicolumn{3}{c}{No MSA} \\
\cmidrule(lr){2-4}\cmidrule(lr){5-7}
Block & Own fold & Partners & Unrelated & Own fold & Partners & Unrelated \\
\midrule
0  & 0.272 & 0.236 & 0.206 & 0.204 & 0.286 & 0.285 \\
9  & 0.555 & 0.320 & 0.145 & 0.243 & 0.239 & 0.219 \\
14 & 0.598 & 0.348 & 0.145 & 0.249 & 0.236 & 0.217 \\
30 & 0.696 & 0.407 & 0.138 & 0.249 & 0.235 & 0.207 \\
40 & 0.777 & 0.468 & 0.142 & 0.264 & 0.247 & 0.216 \\
47 & 0.792 & 0.453 & 0.148 & 0.266 & 0.280 & 0.240 \\
\midrule
Final structures & --- & 0.565 & --- & --- & 0.279 & --- \\
\bottomrule \\
\end{tabular}
\vspace{0.5em}
\caption{\textbf{Sequence-divergent members of the same fold family converge on their shared structure only with the MSA.}
Probe reconstructions for 542 pairs of Sabmark proteins drawn from the same structural group (median 9.4\% sequence identity, 65 aligned residues), decoded from representations generated with or without the MSA, each by the probe trained on representations of the same type. Own fold is the TM-score of each protein's reconstruction against its own final full-MSA structure. Partners and unrelated are TM-scores between two proteins' reconstructions on structurally aligned residues. unrelated pairs come from different groups and are aligned by position. The final row gives the agreement between the partners' final predicted structures with and without the MSA. Values are medians.}
\label{tab:Sabmark}
\end{table}

If the MSA conveys structural rather than strictly sequence-specific information, as the previous results suggest, proteins sharing a fold but little sequence should follow similar trajectories through the Pairformer. 

To test this, we applied the same probes, without retraining, to 98 proteins from Sabmark \cite{10.1093/bioinformatics/bth493}, whose groups comprise structurally related domains of very low sequence identity. Across 542 protein pairs from the same group, with a median sequence identity of 9.4\% and a median of 65 structurally aligned residues, we compared the partners' reconstructions on their aligned residues at every block, both with and without the MSA (Table~\ref{tab:Sabmark}).

With the full MSA from structurally related proteins, the folds became decodable on approximately the same schedule as in the novel cohort, with a median TM-score of 0.56 by block~9 and 0.79 by block~47. More strikingly, the partners began converging on their shared structure early in the pass. By block~9, their reconstructions agreed at a TM-score of 0.32 on the aligned residues, compared with 0.15 for unrelated proteins. By block~47, this agreement reached 0.45, against a ceiling of 0.57 set by the agreement between their final predicted structures. The partners thus recovered a majority of their shared structural similarity within the first ten blocks and roughly 80\% by the end of the first Pairformer pass, while the root-mean-square deviation (RMSD) \cite{Hu2025.09.23.678064} of their aligned residues fell from 5.6 to 3.9~\AA{}.

Crucially, this relative convergence did not depend strongly on sequence identity. Partners with higher sequence identity attained more similar final structures, but when their agreement was expressed as a fraction of that ceiling, convergence was similar across all identity quartiles.

Without the MSA, this pattern was absent. Median TM-scores remained below the fold-decoding threshold, reaching only 0.24 by block~9 and 0.27 by block~47, as in the novel cohort. More importantly, the partners showed little evidence of convergence toward their shared structure. Agreement between their reconstructions on the aligned residues exceeded the unrelated-protein baseline by at most 0.04 throughout the pass, compared with 0.17 at block~9 and 0.31 at block~47 with the MSA. Rather than falling, the RMSD of their aligned residues rose from 6.3 to roughly 8~\AA{}. Even the partners' final no-MSA predictions bore little resemblance to one another, agreeing at an aligned TM-score of only 0.28, approximately half the 0.57 attained with the MSA.

These results suggest that, within the twilight zone of sequence similarity, convergence toward a shared fold is dependent on the shared structural information made accessible by the MSA. Given sequence alone, AlphaFold 3 largely fails to bring these structurally related but sequence-divergent proteins toward a common structure, suggesting that the MSA provides the necessary structural scaffold to identify structurally convergent proteins. 

\subsection{The MSA supplies structure rather than evolutionary relatedness to the query}

\begin{figure}[p]
\centering
\includegraphics[width=\textwidth]{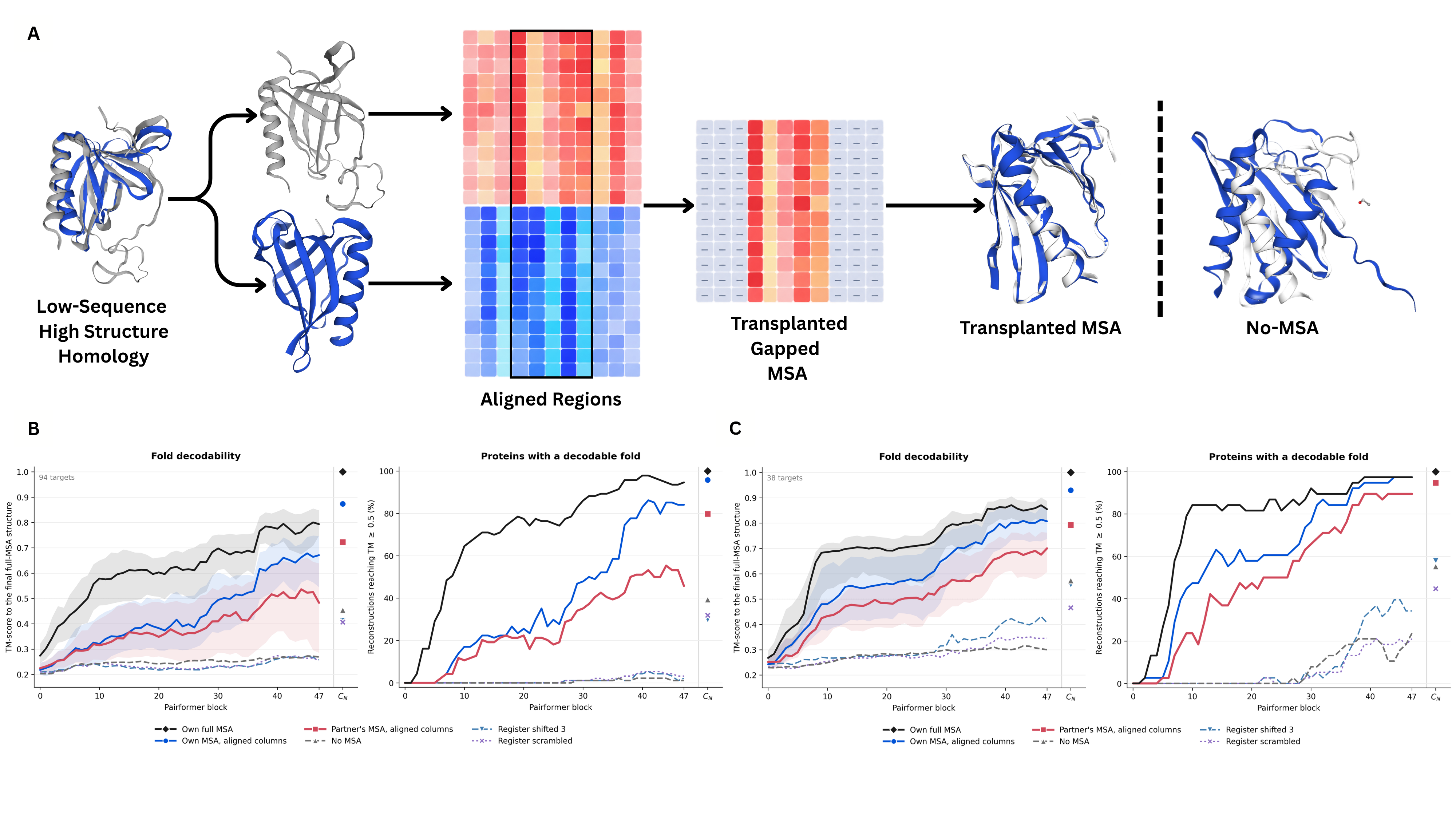}
\caption{\textbf{An alignment built for a different protein recovers the fold, provided it is supplied with the right correspondence.}
\textbf{(A)} Two proteins sharing a fold at low sequence identity are aligned structurally (here the Sabmark domains of 1EVH, in blue, and 1MKE, in grey), and the columns of the donor's alignment that correspond to aligned residues are written onto the target's columns; every other column of the target is gapped. The target's own query sequence is retained as the first row. The resulting prediction is compared against the prediction obtained with no alignment at all, attaining a TM-score of 0.93 against 0.37, so that the transplanted alignment alone rescued an otherwise degenerate fold. In the final overlay, the predicted structure of 1EVH is shown in blue and its experimental structure in white.
\textbf{(B)} Sabmark pairs (94 targets) and \textbf{(C)} novel-cohort pairs (38 targets). Left: median TM-score of the probe reconstruction against the target's own final full-MSA prediction, at each Pairformer block. Right: the percentage of targets whose reconstruction reaches a TM-score of at least 0.5. Lines are medians and shaded bands interquartile ranges; markers to the right of each panel give the diffusion readout of the block-47 representation ($C_N$). Register shifted 3 displaces the transplanted columns three residues along the target chain; register scrambled sends them to a random permutation of the same columns, destroying their order as well as their placement. Both supply the identical donor alignment at the same depth and occupancy. Remaining conditions are reported in Supplementary Tables~S7--S11.}
\label{fig:transplant}
\end{figure}

The preceding sections each revealed something about how AlphaFold 3 converts an alignment into structure, and together they point somewhere specific. With the MSA, the global fold becomes decodable within the first ten blocks of the Pairformer. Which homologs are supplied governs the trajectory of emergence more strongly than which query is supplied, provided the query encodes a compatible structure; indeed, a sequence designed for Q's backbone by inverse folding with ProteinMPNN \cite{Dauparas2022-su}, which never existed in nature and bears no evolutionary relationship to the alignment accompanying it, yields reconstructions more accurate than the native query does. The information is also remarkably sparse, with the MSA-enriched pair representation of only 40 residues recovering a median of 71\% of the accuracy lost when the alignment is removed. And members of the same fold family converge on their shared structure early in the pass, at a rate independent of how much sequence they share, but only when an alignment is present at all.

What these observations suggest is that the value of an alignment may lie not in its evolutionary relationship to the query, but in the structural constraints it encodes. If that is so, then an alignment assembled for an entirely different protein should serve, provided its columns are placed at the structurally equivalent positions of the target. We refer to this correspondence between the donor's alignment columns and the target's residues as the register of the transplanted alignment \cite{Sadreyev2004-xw}. The register is what a structural alignment provides and what sequence similarity would otherwise be used to infer.

We therefore constructed the following test. For two proteins that share a fold at twilight-zone sequence identity, we took every sequence of protein B's alignment and wrote it onto protein A's columns through the structural alignment between them, computed with TM-align \cite{zhang_tm-align_2005}. Only the structurally aligned columns were retained; every other column of A was gapped, so that the transplanted alignment carries no information about the remainder of A's chain. A's own query sequence was left untouched and remains the first row throughout. Three references accompany each transplant: A's own full alignment, which is the ceiling; A's own alignment restricted to exactly the same aligned columns, which controls for the coverage the transplant provides; and no alignment at all, which is the floor (see Methods).

We first applied this to the Sabmark pairs examined in the preceding section: 47 structurally similar pairs, 94 targets in both directions, drawn from the same groups examined above, but with each protein used in at most one pair so that the transplants are independent, sharing a median of 9.4\% sequence identity. The result is remarkable. Without an alignment, these proteins reach a median TM-score of 0.442 against experiment. Given only the structurally aligned columns of a different protein's alignment, with every other column gapped and no sequence in the alignment sharing more than 19.5\% identity with the query, they reach 0.717 ($p = 7\times10^{-9}$). Supplying the target's own alignment over the same columns yields 0.836, and its complete alignment 0.951. Extending the transplant to fill A's unaligned columns with the donor's intervening residues adds little (0.743), confirming that the effect resides in the structurally corresponding columns rather than in the amount of donor sequence supplied. All of this holds with a single diffusion sample and no recycling.

The nulls establish that it is the register, and not the mere presence of a deep alignment, that carries the effect (Figure~\ref{fig:transplant}; Supplementary Tables~S7 and~S9). Each null supplies the same donor alignment at the same depth, occupying the same number of columns, and alters only where those columns land. Sliding the mapping three or seven residues along the target chain yields 0.425 and 0.397; scrambling it, so that the donor's columns land on a random permutation of the target's aligned columns, yields 0.399. All three fall at or below the no-alignment floor, several of them significantly so. A further ablation makes the same point from the opposite direction: overwriting only the structurally aligned columns of the target's own working alignment with residues drawn from a structurally irrelevant region of the donor reduces it from 0.836 to 0.519. Gapping those same columns, rather than corrupting them, yields 0.570, so the loss is not merely the loss of coverage. Those columns are where the structural information resides, and a wrong answer placed in them is worse than no answer at all. Remaining nulls are reported in the Supplementary Information.

Because every Sabmark protein predates AlphaFold 3's training cutoff and may therefore have been encountered during training, we repeated the experiment on proteins that could not have been. From the novel cohort we identified, by all-against-all structural alignment, 19 pairs of proteins sharing a fold (TM-score $\geq 0.5$) at no more than 30\% sequence identity and with substantial $\beta$ content, giving 38 targets (see Methods). The same pattern appears. Without an alignment these proteins reach a TM-score of 0.568; given the partner's aligned columns alone they reach 0.762, against 0.884 for their own alignment over the same columns. The nulls behave as before, with the register-shifted and scrambled conditions at 0.538 and 0.456 and the corrupted-column ablation at 0.479, all at or below the floor. Training exposure therefore does not explain the effect. If anything the novel proteins are easier without an alignment, and the transplant improves them nonetheless.

The emergence trajectories confirm that a transplanted alignment enters the same early pathway as the protein's own. Where the structural alignment covers most of the chain, the transplant is nearly indistinguishable from the target's own alignment through the first twenty-four blocks (0.433 against 0.456 at block~9, and 0.518 against 0.516 at block~24) before falling behind as the pass proceeds. The nulls, by contrast, remain flat from the first block to the last, never rising above the no-alignment trajectory and yielding no decodable fold in any protein before block~27. They do not produce a delayed fold; they produce no early structural signal whatsoever. The benefit of the transplant scales with how much of the target's fold the donor's alignment covers, rising from a median TM-score of 0.567 in the least-covered quartile to 0.856 in the best-covered (Spearman $\rho = +0.51$, $p = 1\times10^{-7}$; Supplementary Table~S11). Sequence identity also correlates with transplant accuracy, but more weakly, and coverage remains the stronger predictor when the two are separated.

We note that the probe reconstructions at block~47 fall well below the diffusion readouts of the same representations (Supplementary Tables~S8--S10). This gap is larger for the transplants than for unmodified alignments, and we attribute it to the probes, which were trained on representations generated from ordinary complete alignments rather than from the heavily gapped and sequence-mismatched alignments used here. The diffusion module, which is the model's own decoder, recovers the fold from these representations without difficulty, and it is the structures it produces that we score against experiment.

The simplest account of these results is that what AlphaFold 3 extracts from an alignment is a description of structure. It does not require sequences related to the query; it requires sequences whose relationship to one another, and to the query's residues, is structural. The alignment need not carry the query's own evolutionary history at all. An alignment assembled for a different family recovers most of the fold, while that same alignment displaced by three residues performs worse than supplying nothing. What the model needs is structural information in the correct register, and when that is supplied, the fold follows.

\section{Discussion}

AlphaFold 3 uses the MSA as a shortcut to predict nonlocal structural features. Without the MSA, the model retains much of the predicted local secondary structure but fails to predict the long-range contacts and partnerships required to organize a correct fold. Restoring the MSA-enriched pair representations of just 40 residues---roughly 10--15\% of a typical chain---recovers a median of 71\% of the lost experimental accuracy. The necessary information in the MSA is therefore distributed, partially redundant, and remarkably sparse. It does not reside in a privileged set of residues, nor does it require the complete MSA-enriched state to attain high recovery.

The Pairformer subsequently reorganizes this MSA-derived contribution. By block 25, a component wholly orthogonal to the original MSA-aligned component dominates the internal representation. Yet this reorganization does not prevent the emergence of the fold; rather, it accompanies it. Probing the pair representation directly shows that, with the MSA, the global fold becomes decodable by block~9 of the 48-block pass in aggregate.

Without the MSA, no protein has an emergent fold by that point, and only 14\% do by the end of the first Pairformer pass. The Pairformer, thus, does not merely propagate the MSA's contribution. It rapidly transforms a compact evolutionary signal into an extractable representation of the predicted global fold, which the remainder of the pass then refines, and which the diffusion module ultimately decodes into coordinates.

The intermediate diffusion readouts show what this transformation looks like to the model's own structural decoder. The earliest diffusion structures are overcompact and featureless, with helices appearing before strands and exact $\beta$-bridge partnerships emerging last---in effect mirroring, within a single forward pass, the same hierarchy of MSA dependence observed when evolutionary information is removed entirely. This compaction, however, belongs to the diffusion model rather than to the representation, as the same pair representations already encode a near-native-sized global fold roughly 27 blocks before the diffusion module can render it.

With the full MSA, the global fold emerges within roughly the first ten blocks. With small, evolutionarily diverse subsets, it takes about thirty-six blocks, even though both arrive at the correct-fold regime by the end of the first pass; the subsets primarily miss some of the finer relationships supplied by the complete alignment. The MSA also influences the initial direction of the trajectory: different sets of homologs generate distinct early emergence pathways much more readily than different query sequences do. This is consistent with previous work showing that MSA subsampling can direct AlphaFold toward alternative conformations in fold-switching proteins, but extends that observation to non-fold-switching proteins and helps localize where in the computation the effect acts. When the query is instead redesigned to suit the backbone, emergence proceeds at much the same pace, but the final structure is more accurate. When the query is shuffled, by contrast, nothing emerges at all. Similarly, as we showed previously, supplying a query with an MSA belonging to a structurally different protein yields an incorrect structure \cite{alphainterp}, meaning the query and its alignment must be compatible for the shortcut to operate.

Taken together, these observations led us to a hypothesis. If the query sequence matters chiefly by encoding a structure, if different sets of homologs separate the trajectory more than different queries do, and if members of a fold family converge only when an alignment is present, then what the model draws from the alignment may not be the evolutionary relationship between those sequences and the query at all, but the structure that relationship happens to encode. If that were so, an alignment assembled for an entirely different protein should serve just as well, provided the two proteins share a fold and the alignment is supplied at the target's structurally corresponding residues.

It does. Supplying a protein with only the structurally aligned columns of a different protein's alignment, with every other column gapped, raises the median TM-score against experiment from 0.442 to 0.717 in Sabmark pairs sharing 9\% sequence identity, and from 0.568 to 0.762 in post-cutoff proteins that could not have been seen during training. The same alignment displaced three residues along the chain, or scrambled across the same columns, performs worse than supplying no alignment at all. Overwriting only those columns in a protein's own working alignment is similarly destructive. 

What the model requires is therefore structural information in the correct register. The alignment is not merely evolutionary evidence from which AlphaFold infers a fold, but a transferable description of structure that the model can read directly when supplied in the correct register.

Our results suggest that the conventional reading of AlphaFold's success, that the model has largely solved structure prediction and that the alignment is a helpful but secondary input, therefore has it backwards. The alignment does not seem to be a hint that the model exploits. Rather, it is where the fold comes from. Without it, the model does not merely do worse---it fails to arrive at a global fold at all within a complete Pairformer pass, and even ten rounds of recycling leave it short of the threshold for a correct fold \cite{alphainterp, feldman_adversarial_2026}. With it, the model need not search. It does not propose an arrangement of secondary structure and test whether the arrangement is consistent. It reads the arrangement from its input and spends the remainder of the computation refining what it was given. Seen this way the behavior is not surprising. A model trained to minimize structural error, given inputs that reliably encode the answer, has no reason to learn to derive from sequence what it can read from an alignment.

This is the titular shortcut: not a shortcut from evolutionary history to structure, but a shortcut from the structure already latent in an alignment to the structure the model outputs. It is extraordinarily effective, and nothing in these results diminishes that. But its effectiveness is also the reason the problem is not closed. Frontier models such as ESMFold2 \cite{Candido2026.06.03.729735} and OpenDDE \cite{project2026foldingreasoningscalingopensource} still perform best when supplied with evolutionary information, which suggests that the shortcut has not been superseded so much as inherited. Until a model can recover protein structure confidently and accurately from sequence alone, without an input that already encodes much of the fold, we do not believe the underlying problem has been solved.

If the model is indeed reading structure from the alignment, the finding also changes what one might do with that fact. Alignment depth has long been treated as a fundamental constraint, and orphan proteins, designed sequences, and fast-evolving families are hard cases because few homologs can be found. Our results suggest the constraint is structural rather than evolutionary, meaning that the provided alignment need not be the protein's own evolutionary history but some other protein's with a similar fold. Where a structural hypothesis exists---from a biophysical argument, a partial model, or a low-resolution experiment---it may be possible to construct an alignment that conveys it. The same logic applies to conformational heterogeneity. MSA subsampling is currently used to coax alternative conformations from the model by removing sequences and observing what changes \cite{cao2026sfclusterfrustrationguidedmsasubsampling, Wayment-Steele2024-rb}. If these results generalize, one might instead supply, at the regions expected to switch, the aligned columns of a protein adopting the alternative fold. We have not attempted either here, and both would require care that our experiments do not supply. But they follow from the mechanism, and they are not available under the view that the alignment's value lies in its evolutionary relationship to the query.

These findings also carry a lesson beyond structure prediction. Predictive accuracy alone does not reveal the computation that produces it, and a model may achieve accurate outputs by exploiting information already present in its inputs rather than by reproducing the process it is used to study. Causal interventions on internal representations, together with readouts of intermediate states, make it possible to distinguish the two, and we expect them to be equally informative for other biological models whose accuracy has outpaced our understanding of how it is achieved.

\newpage

\section{Methods}
\subsection{Protein cohorts}

In this study, we examined two cohorts of 200 monomeric proteins each, assembled in our previous work \cite{alphainterp}. Proteins in the novel cohort had no detectable sequence match exceeding 40\% identity to any sequence deposited in the PDB before September 30, 2021, the cutoff date for structures potentially available during AlphaFold 3 training. Proteins in the similar cohort exceeded this threshold for at least one such sequence. All proteins examined in this study were deposited after the cutoff date. Thus, the two cohorts differ in their degree of sequence similarity to proteins whose structures may have been available during training, allowing us to examine whether this distinction influences AlphaFold 3's predictions (see our previous work for cohort construction).

\subsection{AlphaFold 3 predictions}

All predictions in this study were generated using AlphaFold 3 \cite{abramson_accurate_2024}, with the forward pass modified to allow the extraction of internal representations. We introduced hooks to save these representations at selected checkpoints throughout the model, without modifying the underlying architecture or the computations performed during structure prediction. Full-MSA predictions used the complete alignment generated by the standard AlphaFold 3 data pipeline, while no-MSA predictions used an alignment containing only the query sequence. The number of recycling iterations and diffusion samples varied by experiment,
as specified in the corresponding Methods subsections. To isolate the influence of the MSA alone, no structural templates were used. 

We augmented the forward pass to save the pair representation $P \in \mathbb{R}^{N \times N \times 128}$ and the single representation $S \in \mathbb{R}^{N \times d_s}$ at checkpoints A and B, and following each complete Pairformer pass from $C_1$ through $C_N$. Checkpoint A was extracted following sequence initialization and before the MSA module, while checkpoint B was extracted immediately after the MSA module and before the Pairformer. The Pairformer comprises 48 blocks, numbered 0 through 47 throughout this paper \cite{abramson_accurate_2024}. The checkpoints $C_1$ through $C_N$ refer to representations extracted after complete Pairformer passes, rather than individual blocks. In particular, $C_1$ corresponds to the representation following block~47 of the first Pairformer pass.

In addition, we extracted representations following each of the 48 blocks during the first Pairformer pass for our representational analyses. For the intermediate diffusion experiments, we selected seven of these blocks (14, 18, 24, 30, 36, 40, and 44) and independently generated structural predictions from their saved representations. Together with $C_1$ and $C_N$, this provided nine structural checkpoints per protein under the full-MSA condition. For the corresponding no-MSA analysis, we examined blocks~18, 24, 30, 36, 40, and 44, together with $C_1$ and $C_N$, giving eight checkpoints per protein.

\subsection{Structural comparison metrics}

Structural comparisons involving structures generated by AlphaFold 3 were performed using OpenStructure (OST) \cite{biasini_openstructure_2013}. Unless otherwise stated, the reference was the experimental structure deposited in the PDB; where comparisons were instead made against the full-MSA prediction or another reference structure, this is specified in the corresponding Methods section, Results section, or figure caption. Comparisons involving structures reconstructed from probe-decoded distances, whose residues correspond to the reference by position rather than by sequence alignment, used the TM-score procedure defined in Section~\ref{sec:mds}.

\textbf{Contacts.} A residue pair $(i,j)$ was classified as a contact if its C$\alpha$--C$\alpha$ distance in the reference structure was less than 8~\AA. Pairs separated by fewer than six residues in the sequence were excluded as trivial neighbors \cite{alphainterp, Li2011-wj}. The remaining contacts were partitioned according to their sequence separation into short-range ($6 \leq |i-j| < 12$), medium-range ($12 \leq |i-j| < 24$), and long-range ($|i-j| \geq 24$) contacts. Contact retention was calculated as the fraction of reference contacts preserved in the evaluated prediction. The contact probes described below used C$\beta$ distances, with C$\alpha$ used for glycine, which lacks a C$\beta$ atom. C$\beta$ provides a measure of residue proximity that also incorporates the position of the side chain. Apart from the contact probes and the category-specific recovery analysis, all contact comparisons used C$\alpha$ exclusively.

\textbf{Secondary structure.} Secondary-structure assignments were computed using DSSP \cite{https://doi.org/10.1002/pro.70208, https://doi.org/10.1002/bip.360221211}, implemented through \texttt{mkdssp} \cite{https://doi.org/10.1002/pro.70208}, which was applied independently to each structure. Chain identifiers longer than one character, as produced by AlphaFold 3, were first renamed to single characters. Helices comprised residues assigned H, G, or I, while strands comprised residues assigned E or B \cite{https://doi.org/10.1002/pro.70208, Burley2022-tb}. Retention was calculated as the fraction of residues assigned to a given class in the reference structure that retained the same classification in the evaluated prediction. This is a per-residue comparison and does not require the corresponding structural elements to occupy the same three-dimensional positions.

\textbf{Exact $\beta$-bridge partnerships.} A $\beta$-bridge is a specific hydrogen-bonding pattern between two residues, as defined by DSSP \cite{Bhat2020-sn, https://doi.org/10.1002/pro.70208, Siepen2003-ai}. We identified all bridge partnerships in the reference structure and calculated the fraction for which the same residue pair was also assigned a bridge in the evaluated prediction. This is stricter than strand retention because it requires the recovery of the correct pairing partner, rather than simply the preservation of a strand assignment.

\textbf{Global structural agreement.} We computed a superposition-free geometric agreement metric, defined as the fraction of all upper-triangular C$\alpha$--C$\alpha$ pairwise distances in the evaluated structure that differed from their corresponding distances in the reference structure by no more than 2.0~\AA. This measures the preservation of three-dimensional relationships between residue pairs without requiring structural superposition or secondary-structure assignment. Unlike TM-score, which summarizes structural similarity following alignment, this metric directly measures the fraction of pairwise distances that agree within a specified tolerance.

\textbf{Radius of gyration.} We computed the radius of gyration from C$\alpha$ coordinates and reported it as a ratio relative to the final full-MSA prediction, $C_N$. Values below 1.0 indicate that the evaluated structure is more compact than the final full-MSA prediction.

\subsection{Activation patching at checkpoint B}
\label{sec:patching}

To test whether partial restoration of the MSA-enriched pair representation could recover lost structural relationships, we performed activation patching at checkpoint B, immediately after the MSA module but before the Pairformer. For each protein, we obtained the pair representations from both a full-MSA and a no-MSA forward pass, denoted $P_B^{\mathrm{MSA}}$ and $P_B^{\mathrm{noMSA}}$, respectively. We then defined the MSA-enriched difference as

\[
\Delta_B = P_B^{\mathrm{MSA}} - P_B^{\mathrm{noMSA}}.
\]

Given a set of selected anchor residues, $\mathcal{A}$, we constructed a patched representation by replacing the rows and columns corresponding to these residues in the no-MSA representation with their full-MSA values:

\[
{
\widetilde{P}_{B,ij} =
\begin{cases}
P_{B,ij}^{\mathrm{MSA}},
& \text{if } i \in \mathcal{A} \text{ or } j \in \mathcal{A}, \\[6pt]
P_{B,ij}^{\mathrm{noMSA}},
& \text{otherwise}.
\end{cases}
}
\]

This patched representation was then passed through the Pairformer and diffusion module to generate a predicted structure. Because selecting a single anchor restores its pairwise relationships with every other residue in the protein, our intervention reintroduces the broader representational context of that residue, rather than simply restoring its isolated local properties.

\subsection{Anchor selection strategies}

Anchors were selected according to one of three strategies:

\textbf{Contact-selected.} Residues were ranked by the number of long-range contacts in which they participated within the full-MSA prediction, and the top $k$ residues were selected, with ties broken arbitrarily.

\textbf{Conservation-selected.} Residues were ranked by increasing column entropy in the MSA, with lower entropy indicating greater sequence conservation, and the top $k$ residues were selected \cite{shannon_mathematical_nodate}.

\textbf{Random.} Residues were selected uniformly at random without replacement.

For the disjoint-set analysis (Supplementary Figure~S1D), we partitioned each ranked list into four mutually exclusive sets of 40 residues and performed the patching experiment independently for each set. The titration across anchor counts and selection strategies is described in Supplementary Methods, Section~S8.

\subsection{MSA-module contribution}

We isolated the MSA-specific contribution of the MSA module by calculating the change in the pair representation between checkpoints A and B under both the full-MSA and no-MSA conditions:

\[
\begin{aligned}
U_{\mathrm{MSA}}
    &= P_B^{\mathrm{MSA}} - P_A^{\mathrm{MSA}}, \\[6pt]
U_{\mathrm{noMSA}}
    &= P_B^{\mathrm{noMSA}} - P_A^{\mathrm{noMSA}}.
\end{aligned}
\]

We then defined the MSA-specific module contribution as the difference between these two updates:

\[
M = U_{\mathrm{MSA}} - U_{\mathrm{noMSA}}.
\]

Writing the difference already present at checkpoint A as

\[
\Delta_A = P_A^{\mathrm{MSA}} - P_A^{\mathrm{noMSA}},
\]

we obtain the exact identity

\[
{\Delta_B = \Delta_A + M.}
\]

This identity was verified numerically for every protein, with a residual below $10^{-6}$.

To determine how much of the difference at checkpoint B was attributable to the MSA-specific module contribution, we calculated the projection coefficient:

\[
\frac{\langle M,\Delta_B\rangle}
     {\|\Delta_B\|_F^2}.
\]

The reported contribution of 99.5\% corresponds to this coefficient. We also calculated the corresponding coefficient for $\Delta_A$ and the cosine similarity between $M$ and $\Delta_A$, after flattening the representations into vectors.

\subsection{Magnitude--direction decomposition}

For each residue pair $(i,j)$, we decomposed the 128-dimensional MSA-specific module contribution, $M_{ij}$, into its magnitude and unit direction:

\[
\begin{aligned}
m_{ij} &= \|M_{ij}\|_2, \\[6pt]
\widehat{M}_{ij} &= \frac{M_{ij}}{\|M_{ij}\|_2}.
\end{aligned}
\]

We used this decomposition to characterize the spatial organization of the MSA-specific contribution. We also used the same magnitude--direction decomposition for the swap experiments described below, although those experiments decomposed the native and shuffled module updates separately rather than decomposing their difference, $M$.

To examine how the magnitude was distributed across residue pairs, we formed an $N \times N$ matrix containing the magnitude of each pairwise contribution. We then computed its singular-value decomposition and calculated the effective rank \cite{roy_effective_2007}:

\[
\operatorname{erank} = \exp(H),
\]

where

\[
H = -\sum_k p_k\log p_k,
\qquad
p_k = \frac{\sigma_k^2}{\sum_j \sigma_j^2}.
\]

Here, $\sigma_k$ denotes the $k$th singular value, and $p_k$ is its normalized contribution to the squared singular-value spectrum.

\subsection{Within-column MSA shuffling}
\label{sec:shuffling}

To disrupt cross-column evolutionary relationships while preserving the amino-acid composition of individual alignment columns, we independently shuffled selected MSA columns, following the protocol described in our previous study \cite{alphainterp}. Within each selected column, we randomly permuted the amino-acid identities of homologous sequences while keeping the query sequence and all gap positions fixed. This preserved the amino-acid frequencies and gap pattern of each column while disrupting relationships between amino acids occurring together across different columns.

We applied this perturbation at seven shuffle fractions: 10\%, 20\%, 40\%, 70\%, 90\%, 95\%, and 99\% of the eligible columns. At each fraction, we generated a new AlphaFold 3 prediction and extracted the corresponding MSA-specific module contribution, $M'$.

To quantify the representational changes caused by shuffling, we calculated the cosine similarity and relative perturbation between the native and shuffled contributions:

\[
\cos(M,M')
=
\frac{\langle M,M'\rangle}
{\|M\|_F\|M'\|_F},
\]

\[
\text{Relative perturbation}
=
\frac{\|M-M'\|_F}{\|M\|_F}.
\]

These calculations were performed after flattening the representations into vectors. We also measured the Jaccard index between the C$\alpha$-based contact maps of the native and shuffled structural predictions.

\subsection{Magnitude--direction swap experiments}

To determine whether the structural consequences of the MSA-module contribution depended on its magnitude, its direction, or both, we calculated the module updates under the native and 99\%-shuffled conditions:

\[
\begin{aligned}
U_c &= P_B^c - P_A^c, \\[6pt]
U_s &= P_B^s - P_A^s.
\end{aligned}
\]

Here, $c$ denotes the native alignment and $s$ denotes the shuffled alignment. These are the module updates within each condition, rather than the MSA-specific difference $M$ defined above.

We independently decomposed these two updates into magnitude and direction for each residue pair and constructed four conditions by recombining their corresponding components. For each residue pair $(i,j)$, we defined

\[
\begin{aligned}
\mathrm{cc:}\quad
&\|U_{c,ij}\|_2\widehat{U}_{c,ij},
&&\text{native magnitude, native direction}, \\[6pt]
\mathrm{sc:}\quad
&\|U_{s,ij}\|_2\widehat{U}_{c,ij},
&&\text{shuffled magnitude, native direction}, \\[6pt]
\mathrm{cs:}\quad
&\|U_{c,ij}\|_2\widehat{U}_{s,ij},
&&\text{native magnitude, shuffled direction}, \\[6pt]
\mathrm{ss:}\quad
&\|U_{s,ij}\|_2\widehat{U}_{s,ij},
&&\text{shuffled magnitude, shuffled direction}.
\end{aligned}
\]

For each hybrid update, $U_h$, we constructed the target checkpoint-B representation by adding the hybrid to the native checkpoint-A representation:

\[
P_B^{\mathrm{target}}
=
P_A^c + U_h.
\]

Because computation proceeded using a no-MSA forward pass, the live representation at checkpoint B was initially $P_B^{\mathrm{noMSA}}$. We therefore added the following increment to the live representation:

\[
\Delta_{\mathrm{inject}}
=
P_B^{\mathrm{target}}
-
P_B^{\mathrm{noMSA}}.
\]

This procedure replaces the live checkpoint-B state with the intended hybrid representation. Under the control condition, the target becomes

\[
P_A^c + U_c
=
P_A^c + (P_B^c-P_A^c)
=
P_B^c,
\]

exactly reproducing the native checkpoint-B representation. Computation then proceeded normally through the Pairformer and diffusion module, and the resulting structures were compared against the full-MSA prediction using TM-score.

For each protein, we defined the direction rescue as

\[
R_{\mathrm{direction}}
=
\frac{T_{\mathrm{sc}}-T_{\mathrm{ss}}}
     {T_{\mathrm{cc}}-T_{\mathrm{ss}}},
\]

and the magnitude rescue as

\[
R_{\mathrm{magnitude}}
=
\frac{T_{\mathrm{cs}}-T_{\mathrm{ss}}}
     {T_{\mathrm{cc}}-T_{\mathrm{ss}}},
\]

where $T$ denotes the TM-score against the full-MSA prediction.

A rescue value of 1.0 indicates that restoring the corresponding component recovers the control structural prediction to the same extent as the native contribution, while a value of 0.0 indicates no recovery relative to the shuffled condition.

To avoid dividing by a near-zero denominator, we required

\[
T_{\mathrm{cc}}-T_{\mathrm{ss}}>0.1.
\]

Of the 200 novel proteins examined, 43 did not pass this filter and were excluded from the rescue-ratio analysis.

\subsection{MSA-module persistence across Pairformer blocks}

To track how much of the original MSA-specific module contribution remained aligned with the evolving pair representation, we projected the difference between the full-MSA and no-MSA representations at each Pairformer block onto the original contribution, $M$.

For each block $l$, we defined the MSA-dependent component as

\[
\widetilde{P}^{(l)}
=
P_{\mathrm{MSA}}^{(l)}
-
P_{\mathrm{noMSA}}^{(l)}.
\]

Both representations were extracted at the same block, allowing us to isolate the difference between the two conditions at that depth. We then calculated the projection coefficient

\[
{
\operatorname{proj}_M(l)
=
\frac{\langle \widetilde{P}^{(l)},M\rangle}
{\|M\|_F^2}.
}
\]

This scalar measures how much of the original contribution remains aligned with the evolving representation. These values are projection coefficients, rather than normalized component magnitudes.

We further decomposed the representation into the component aligned with $M$ and an orthogonal residual:

\[
\begin{aligned}
P_M^{(l)}
&=
\operatorname{proj}_M(l)M, \\[6pt]
R^{(l)}
&=
\widetilde{P}^{(l)}-P_M^{(l)}.
\end{aligned}
\]

To compare the relative magnitudes of these components, we calculated

\[
\frac{\|P_M^{(l)}\|_F}
     {\|\widetilde{P}^{(l)}\|_F}
\qquad\text{and}\qquad
\frac{\|R^{(l)}\|_F}
     {\|\widetilde{P}^{(l)}\|_F}.
\]

These normalized component magnitudes are distinct from the projection coefficient itself.

\subsection{Contact probes across Pairformer blocks}
\label{sec:probes}

To determine when contact information became decodable from the pair representation, we trained ridge-regression \cite{hoerl_ridge_nodate} probes and evaluated them at each of the 48 Pairformer blocks and checkpoint B, giving 49 probes in total.

\textbf{Training.} Each probe was a standardized ridge-regression classifier operating on the 128-dimensional pairwise feature vector $P_{ij}$. Features were standardized to zero mean and unit variance, and the regression was fitted with a regularization parameter of $\lambda=10.0$.

We assigned a binary target of $+1$ to residue pairs with a C$\beta$--C$\beta$ distance below 8~\AA{}, using C$\alpha$ for glycine, and $-1$ to pairs with a distance of at least 10~\AA{}. Pairs between 8 and 10~\AA{} were excluded to separate the positive and negative classes more clearly. Only long-range residue pairs, with sequence separation of at least 24 residues, were included.

For probe training, we used 100 similar-cohort proteins and reserved another 40 for generic evaluation. All 200 novel-cohort proteins were excluded from both sets and reserved for the shuffling and hybrid analyses. The three groups were mutually disjoint, and the same trained probe weights were used for all subsequent evaluations.

Performance was measured using the area under the receiver operating characteristic curve (AUROC) \cite{Fawcett2006-wf}.

\textbf{Native-versus-shuffled contact preference.} We then applied the trained probes, without refitting them, to the four magnitude--direction conditions at each Pairformer block.

For each protein, we identified native-only contacts, which were present in the native prediction but absent from the shuffled prediction, and shuffled-only contacts, which were present in the shuffled prediction but absent from the native prediction.

We measured how frequently the probe assigned a higher contact score to a native-only pair than to a shuffled-only pair, matching the compared pairs by sequence separation within a tolerance of four residues.

We refer to this separation-matched ranking statistic as the control preference. A value of 0.5 corresponds to chance-level discrimination, while a value of 1.0 indicates that native-only contacts were always ranked higher. Ties were assigned a value of 0.5.

\textbf{Generic versus fold-specific contact evaluation.} Generic AUROC measured contact versus non-contact discrimination on the 40 held-out similar-cohort proteins. Fold-specific AUROC measured discrimination between contacts unique to the native and shuffled predictions on the 200 reserved novel-cohort proteins, using the same frozen probe weights and sequence-separation-matched comparisons. The hybrid-preference analysis also used these 200 novel proteins.

\subsection{Intermediate diffusion readouts}
\label{sec:emergence}

To determine when the evolving pair representation could support accurate coordinate generation, we generated independent structural predictions from saved intermediate Pairformer states.

At each checkpoint, we passed the saved pair and single representations to the unmodified diffusion module and generated a single diffusion sample. Under the full-MSA condition, we examined seven Pairformer blocks (14, 18, 24, 30, 36, 40, and 44), together with $C_1$ and $C_N$. This provided nine structural checkpoints for each of the 400 proteins.

Under the no-MSA condition, we examined the corresponding checkpoints except for block~14, providing eight structural checkpoints for the 200 novel proteins.

Although these representations were collected from successive stages of the same Pairformer computation, the resulting structures were generated through separate diffusion runs. They therefore represent independent structural readouts of the evolving representation, rather than successive coordinate states from a single diffusion trajectory.

Each intermediate structure was compared against the experimental structure using TM-score. We also compared the intermediate structures against the final full-MSA prediction using the structural metrics described above.

Because the diffusion module was trained to generate coordinates from the completed representation, its behavior on intermediate representations may differ from its behavior during a standard prediction. These intermediate structures should therefore be interpreted as readouts of the structural information accessible to the diffusion module at each checkpoint, rather than as optimized structural predictions or physical folding intermediates.

\subsection{Disjoint-MSA subsets}
\label{sec:disjoint}

To test whether the structural trajectory through the Pairformer depended on which homologs were provided, we constructed two evolutionarily independent MSA subsets for each qualifying protein in the novel cohort.

From each protein's full MSA, we pooled every homolog sharing $\leq 30\%$ identity with the query and covering $\geq 50\%$ of the query-aligned columns. Identity between any two sequences was computed over query-aligned columns, excluding positions where either sequence carried a gap; pairs sharing fewer than 20\% of query columns were treated as unverifiable and rejected. From this pool we drew two disjoint subsets of five sequences each by greedy selection from a random permutation, adding a candidate to subset A only if it shared $\leq 50\%$ identity with every sequence already in A, and to subset B only if it additionally shared $\leq 30\%$ identity with every sequence in A. Selection was restarted with a new permutation if either subset could not be completed. Of the 200 novel proteins, 100 satisfied all constraints; across these, the maximum cross-subset identity was 0.300 and the median 0.289.

Each protein was folded twice---once with each subset---using 0 recycles, a single diffusion sample, no templates, and intermediate saving at all 48 Pairformer blocks, with the model seed held fixed. Intermediate structures were generated from blocks 14, 18, 24, 30, 36, 40, 44, and the final output ($C_N$) using the diffusion protocol described above. 

\subsection{Query substitution}
\label{sec:queries}

To separate the contribution of the query sequence from that of the alignment, we replaced the query while holding the subsets fixed. Each variant was folded with both subsets under the settings above, giving four conditions per subset: the original query (Q), a natural homolog (H), a designed sequence (MPNN), and a shuffled sequence.

\textbf{Homolog queries (H).} From each protein's full MSA we selected a homolog sharing 30--90\% identity with the query and covering $\geq 90\%$ of its columns, excluding any sequence belonging to subset A or B, and preferring homologs with an experimental structure in the PDB (at least 95\% identity over at least 90\% of their length); among the remainder we took the candidate closest to 60\% identity. Each subset was then re-anchored onto the homolog: using the homolog's own row in the query's alignment, columns occupied by the homolog were retained as match states, columns where it carried a gap became insertions, and its own insertions became new columns. Every re-anchored row was verified to preserve its residues exactly and to contain one match state per homolog residue. Of the 100 proteins, 96 had an eligible homolog; the median identity to the query was 0.60.

\textbf{Designed queries (MPNN).} We used ProteinMPNN \cite{Dauparas2022-su} to design a new sequence for each protein's own predicted backbone, taken from its final full-MSA prediction. We generated one sequence per backbone at a sampling temperature of 0.3 with a fixed random seed, with no positions fixed, no residues omitted, and no bias applied. The resulting designs shared a median of 45\% identity with the native query (range 25--57\%) and, by construction, matched its length exactly.

\textbf{Shuffled queries.} We randomly permuted the residues of the query sequence, preserving its length and amino-acid composition while destroying any structural signal it carried. The median identity to the native query was 5\%.

Because the designed and shuffled queries preserve the length of the original, they occupy the same alignment columns, and only the query row of each subset was replaced. For the homolog queries, residues were mapped back to the original numbering through the re-anchoring described above.

\subsection{Probing the pair representation}
\label{sec:probe}

To determine what structural information the pair representation contains, independently of the diffusion module, we trained linear probes to decode final C$\alpha$--C$\alpha$ distances from the pair representation at each stage of the first Pairformer pass. We used the representations at checkpoints A and B and after each of the 48 Pairformer blocks, giving 50 layers per protein, from runs with the full MSA and from runs with the query sequence alone.

For each unordered residue pair $(i,j)$ at layer $\ell$, the input was the concatenation of the two directional pair embeddings and a constant term,

\[
x^{(\ell)}_{ij} = \bigl[\,P^{(\ell)}_{ij} \,\|\, P^{(\ell)}_{ji} \,\|\, 1\,\bigr] \in \mathbb{R}^{257},
\]

and the target was the C$\alpha$--C$\alpha$ distance $y_{ij}$ between residues $i$ and $j$ in the protein's final full-MSA prediction. The target was therefore held fixed across layers, and only the representation varied. A separate ridge-regression probe was fitted at each layer by minimizing a protein-equal-weighted squared error,

\[
\mathcal{L} = \frac{1}{P}\sum_{p=1}^{P} \frac{1}{n_p}\sum_{(i,j)\in p}\bigl(y_{pij} - x_{pij}^{\top}\beta\bigr)^2 + \lambda\|\beta\|^2,
\qquad \lambda = 1,
\]

where $P$ is the number of training proteins and $n_p$ is the number of residue pairs in protein $p$, so that longer proteins did not dominate the fit. Probes were trained exclusively on the 200 proteins of the similar cohort and evaluated exclusively on the 200 proteins of the novel cohort, so that no evaluated protein was seen during training. Two families of probes were trained: one on representations generated with the full MSA, and one on representations generated without it. Each condition was decoded by the probe trained on representations of the same type. As a control for differences in representational format, each probe was also applied to representations of the other type. Negative decoded distances were set to zero. The same full-MSA probe was applied to the disjoint-subset and query-substitution runs.

\subsection{Reconstruction and evaluation of decoded structures}
\label{sec:mds}

For each protein and layer, the decoded distances were assembled into a symmetric matrix $\hat{D}$ and converted into three-dimensional C$\alpha$ coordinates by classical multidimensional scaling \cite{Sims2005-fq}. The squared distance matrix was double-centered,

\[
B = -\tfrac{1}{2}\, J \hat{D}^{(2)} J,
\]

where $J$ is the centering matrix, and coordinates were obtained from the eigenvectors of the three largest eigenvalues of $B$, scaled by the square roots of those eigenvalues. Because classical multidimensional scaling determines coordinates only up to reflection, both the reconstruction and its mirror image were evaluated, and the higher score was retained.

Each reconstruction was compared against the final full-MSA prediction using TM-score with the residue correspondence fixed by sequence position,
\[
\mathrm{TM} = \frac{1}{L}\sum_{i=1}^{L}\frac{1}{1 + (d_i/d_0)^2}, \qquad d_0 = 1.24\,(L-15)^{1/3} - 1.8,
\]
where $d_i$ is the distance between corresponding C$\alpha$ atoms after superposition and $L$ is the number of residues. The superposition was chosen to maximize TM-score, searching over superpositions seeded from contiguous fragments of the chain and refined iteratively.

We additionally computed four measures directly from the decoded distances or reconstructions: the coefficient of determination $R^2$ between decoded and final distances over all residue pairs; per-residue lDDT computed from the decoded distances, defined for residue $i$ as the fraction of partners $j$ within 15~\AA{} in the final structure whose decoded distance lay within 0.5, 1, 2, and 4~\AA{} of the final distance, averaged over the four thresholds, and calculated both over all partners and over long-range partners alone ($|i-j| \geq 24$); the ratio of the reconstruction's radius of gyration to that of the final structure; and the Jaccard index between the long-range contact maps (C$\alpha$--C$\alpha$ distance below 8~\AA{}, $|i-j| \geq 24$) of the reconstruction and the final structure. A protein was considered to have a decodable fold at a given layer if its reconstruction reached a TM-score of at least 0.5 \cite{zhang_tm-align_2005}. All measures were computed per protein and reported as medians across the novel cohort, with interquartile ranges. Control baselines are described in Supplementary Section~S5.

For comparison with the diffusion readouts (Section~\ref{sec:emergence}), the readouts generated with and without the MSA were evaluated against the same final full-MSA prediction, using the same TM-score procedure.

\subsection{Trajectory comparison}
\label{sec:disjoint_analyses}

Trajectories were compared in two ways. Progress toward the fold was the TM-score of each run's probe reconstruction against the original query's final full-MSA prediction, computed as described above; for homolog queries, residues were first mapped back to the original numbering, and the comparison was restricted to mapped positions. Similarity between runs was the TM-score between the probe reconstructions of two runs of the same protein at the same layer, with residue correspondence fixed by position, taking the mean of the two normalizations and retaining the better of the two mirror images. Runs differing in query were compared within the same MSA subset.

\subsection{Display of reconstructed structures}
\label{sec:display}

For the structures shown in Figure~\ref{fig:probe}B--C, each classical reconstruction was refined against its decoded distances to improve local geometry for display. Coordinates were optimized by minimizing the weighted squared difference between reconstructed and decoded distances, with weights $1/\max(\hat{D}_{ij}, 3.8)^2$ so that short distances were fitted more closely, together with a harmonic restraint holding consecutive C$\alpha$ atoms at 3.8~\AA{} and a penalty on any C$\alpha$ pair closer than 3.8~\AA{}. Optimization used L-BFGS for up to 300 iterations. Each refined reconstruction, and its mirror image, was superimposed onto the final structure by the TM-score-maximizing superposition described above, and the orientation with the higher TM-score was retained. Full backbones were then rebuilt from the C$\alpha$ traces with PULCHRA \cite{Rotkiewicz2008-tp} for visualization. Residues were colored by the distance between their reconstructed and final C$\alpha$ positions. This refinement was used only for display; all reported measures were computed from the unrefined classical reconstructions.

\subsection{MSA transplantation}
\label{sec:transplant}

To test whether the structural information AlphaFold 3 extracts from an alignment depends on that alignment's evolutionary relationship to the query, we supplied each target protein with the alignment of a different protein that shares its fold.

\textbf{Pair selection.} For Sabmark, structural alignments between domains of the same group were taken from the benchmark's own correspondences. Pairs were required to share at least 30 aligned residues, and each protein was used in at most one pair. From the same 98 proteins examined above, this gave 47 pairs and 94 ordered transplants, each protein serving once as target and once as donor. No pair exceeds 19.5\% sequence identity and the median is 9.4\%, so no donor could be retrieved by a sequence search starting from its target. We verified this directly: across all 47 pairs, not one sequence appears in both the donor's and the target's alignment, from alignments averaging roughly 2{,}000 sequences each, and the highest identity between a donor's query and any sequence in its target's alignment is 27.3\% in the median, against 25.2\% for proteins drawn from different structural groups. By this measure the pairs are indistinguishable from unrelated proteins, and whatever structural information a transplant supplies was unavailable to the target through its own homology search. Classified against SCOP 1.65, the release on which Sabmark is built, 32 of the 47 pairs fall in different families and 15 of those in different superfamilies, where the superfamily boundary reflects a judgement about the evidence for common descent rather than a fixed threshold, so separation at that level does not exclude remote homology, and our interpretation does not rest on it.

For the novel cohort, no such correspondences exist, so we computed all pairwise structural alignments among the 200 final full-MSA predictions with TM-align \cite{zhang_tm-align_2005} and took the residue correspondence from the resulting alignment. We retained pairs with a TM-score of at least 0.5 normalized by the shorter chain, no more than 30\% sequence identity over the aligned residues, alignment coverage of at least 0.5 in both partners, and a DSSP strand fraction of at least 0.15 in both, so that the folds examined were not helical bundles. Using each protein once gave 19 pairs and 38 ordered transplants, spanning 6.7--23.7\% sequence identity and TM-scores of 0.50--0.69. To our knowledge, these proteins postdate SCOPe's coverage and are unclassified, so their pairs were selected on structural similarity and sequence divergence alone, without reference to any assertion about their evolutionary relationship. Alignment coverage of a target is the number of aligned residues divided by the length of that target's chain; for pair-level filters we used the smaller of the two partners' coverages.

\textbf{Transplant construction.} Let $(i,j)$ denote the residue pairs of the structural alignment between target A and donor B. Every sequence of B's unpaired alignment was rewritten onto A's columns by placing the residue at B's column $j$ into A's column $i$; insertions, which occupy no column of A, were discarded. In the gapped transplant, every other column of A was set to a gap, so that the transplanted alignment describes only those positions whose structural correspondence is established. In the ungapped transplant, A's unaligned columns were additionally filled from B's intervening residues, with any surplus trimmed from the middle of each stretch. A's own query sequence was retained unchanged as the first row of every condition, and the paired alignment was left empty throughout, including in the controls, so that all conditions rely on the unpaired alignment alone. Rows retaining fewer than ten residues after mapping were discarded.

\textbf{Controls and ablations.} The coverage-matched control retains A's own alignment but gaps every column outside the aligned set, so that it supplies the same positions as the transplant from the target's own homologs. The register-shift ablations displace the mapping $k$ residues along A ($i \rightarrow i+k$, $k = 3, 7$), preserving donor residue adjacency while destroying structural correspondence; the scramble ablation sends the donor's columns to a random permutation of A's aligned columns, destroying both. The corrupted-column ablation retains A's own alignment everywhere except the aligned columns, which are overwritten row by row with residues drawn from structurally unaligned positions of the donor; the masked ablation gaps those same columns instead, separating the loss of correct information from the introduction of incorrect information. The donor-complement ablation supplies only the donor content falling outside the aligned columns. Each ablation supplies the same donor alignment at the same depth, and the register-shift and scramble ablations occupy the same number of columns as the transplant itself, so that they differ from it only in where those columns land.

\textbf{Prediction and scoring.} All transplants, controls and ablations were folded with 0 recycles, a single diffusion sample, no templates, a fixed model seed, and intermediate saving at all 48 Pairformer blocks. Final structures were compared against the experimental structure, taking the chain named by the domain identifier for Sabmark domains and the longest chain for the novel cohort, with TM-score normalized by the predicted structure so that a domain is scored within its experimental chain; they were also compared against the target's own final full-MSA prediction. Intermediate representations were decoded with the full-MSA distance probes of Section~\ref{sec:probe}, without retraining, and reconstructed as described in Section~\ref{sec:mds}.

\textbf{Contiguous-block transplant.} As a variant of the gapped transplant, we copied only runs of at least three consecutive aligned residue pairs, discarding fragmented single-residue correspondences and preserving the donor's own residue adjacency within each block. On the higher-coverage Sabmark pairs this produced alignments identical to the gapped transplant, since those structural alignments contain no runs shorter than three residues, and it was therefore not extended to the remaining pairs.

\subsection{Sabmark trajectories}
\label{sec:Sabmark}

We applied the distance probes described above (Section~\ref{sec:probe}), without retraining, to the 48-block pair representations of 98 Sabmark \cite{10.1093/bioinformatics/bth493} proteins folded with and without their full MSAs, decoding each condition with the probe trained on representations of the same type and using each protein's final full-MSA prediction as its reference. Residue correspondence between proteins of the same group was taken from their structural alignments. For each same-group pair with at least 30 aligned residues, we computed at every block the TM-score between the two reconstructions on their aligned residues, taking the mean of the two normalizations and retaining the better of the two mirror images, together with the C$\alpha$ RMSD of the aligned residues after optimal superposition. As a ceiling, the same measures were computed between the partners' final predicted structures, with and without the MSA. As a baseline, each protein was paired with a protein from a different group, and the two were aligned by position over the same number of residues. To relate convergence to sequence identity, we divided pairs into identity quartiles and compared both the raw agreement and the agreement expressed as a fraction of the full-MSA ceiling.

\subsection{Recycling}

Except where stated, all predictions in this study used a single Pairformer pass with no recycling and one diffusion sample, so that each analysis reflects a single forward computation. The exception is the no-MSA condition in the novel cohort, for which we additionally generated structures at every recycling iteration from $C_1$ through $C_{10}$, allowing us to establish how far repeated passes carry a prediction made without evolutionary information.

\subsection{Statistical reporting}

All reported medians, fractions, and counts were calculated across the relevant set of proteins: the 200 novel proteins, the 200 similar proteins, the 400 proteins pooled, the 100 novel proteins used in the disjoint-subset and query-substitution analyses (96 for homolog queries), the 94 Sabmark targets and 542 same-group pairs used in the fold-partner and transplant analyses, or the 38 novel-cohort targets of the transplant replication, as specified in each analysis. Measures from the two MSA subsets were pooled when reporting medians; for paired comparisons, values were averaged over the two subsets within each protein. Where a condition was run on only part of a set, the number of proteins contributing is given with the result. Shaded bands in figures---whether in the manuscript or Supplementary Information---denote interquartile ranges across proteins.

We used Wilcoxon signed-rank tests \cite{Wilcoxon1945-jl} for paired comparisons between conditions within the same proteins. All statistical tests were two-sided, and the corresponding $p$-values are reported with the relevant results.

\section{Acknowledgments}

This research was supported in part by a gift from the Parker H. Petit AI-Driven Drug Discovery Initiative and grant GM-118039 from the National Institute of General Medical Sciences of the National Institutes of Health.

We thank Jessica Forness for
proofreading and polishing this manuscript. We would also like to thank Ishan Khire, Giri Krishnan, Tyler Hayes, and the rest of the VizFold team, whose methodology for predicting three-dimensional structure from the pair representations inspired ours.

\bibliographystyle{unsrt}
\bibliography{Shakhnovich}

@article{mkdssp,
	title        = {DSSP 4: FAIR annotation of protein secondary structure},
	author       = {Hekkelman, Maarten L. and Salmoral, Daniel Álvarez and Perrakis, Anastassis and Joosten, Robbie P.},
	year         = 2025,
	journal      = {Protein Science},
	volume       = 34,
	number       = 8,
	pages        = {e70208},
	doi          = {https://doi.org/10.1002/pro.70208},
	url          = {https://onlinelibrary.wiley.com/doi/abs/10.1002/pro.70208},
	eprint       = {https://onlinelibrary.wiley.com/doi/pdf/10.1002/pro.70208}
}

@article{Wayment-Steele2024-rb,
	title        = {Predicting multiple conformations via sequence clustering and {AlphaFold2}},
	author       = {Wayment-Steele, Hannah K and Ojoawo, Adedolapo and Otten, Renee and Apitz, Julia M and Pitsawong, Warintra and H{\"o}mberger, Marc and Ovchinnikov, Sergey and Colwell, Lucy and Kern, Dorothee},
	year         = 2024,
	month        = jan,
	journal      = {Nature},
	publisher    = {Springer Science and Business Media LLC},
	volume       = 625,
	number       = 7996,
	pages        = {832--839},
	copyright    = {https://creativecommons.org/licenses/by/4.0},
	language     = {en}
}

@article{Sadreyev2004-xw,
	title        = {Estimates of statistical significance for comparison of individual positions in multiple sequence alignments},
	author       = {Sadreyev, Ruslan I and Grishin, Nick V},
	year         = 2004,
	month        = aug,
	journal      = {BMC Bioinformatics},
	publisher    = {Springer Science and Business Media LLC},
	volume       = 5,
	number       = 1,
	pages        = 106,
	copyright    = {http://creativecommons.org/licenses/by/2.0},
	language     = {en}
}

@article{Burley2022-tb,
	title        = {Protein Data Bank: A comprehensive review of {3D} structure holdings and worldwide utilization by researchers, educators, and students},
	author       = {Burley, Stephen K and Berman, Helen M and Duarte, Jose M and Feng, Zukang and Flatt, Justin W and Hudson, Brian P and Lowe, Robert and Peisach, Ezra and Piehl, Dennis W and Rose, Yana and Sali, Andrej and Sekharan, Monica and Shao, Chenghua and Vallat, Brinda and Voigt, Maria and Westbrook, John D and Young, Jasmine Y and Zardecki, Christine},
	year         = 2022,
	month        = oct,
	journal      = {Biomolecules},
	publisher    = {MDPI AG},
	volume       = 12,
	number       = 10,
	pages        = 1425,
	copyright    = {https://creativecommons.org/licenses/by/4.0/},
	language     = {en}
}

@article{Bhat2020-sn,
	title        = {{$\beta$-Strand-mediated} interactions of protein domains},
	author       = {Bhat, Archana S and Kinch, Lisa N and Grishin, Nick V},
	year         = 2020,
	month        = nov,
	journal      = {Proteins},
	publisher    = {Wiley},
	volume       = 88,
	number       = 11,
	pages        = {1513--1527},
	copyright    = {http://onlinelibrary.wiley.com/termsAndConditions\#vor},
	language     = {en}
}

@article{alphainterp,
	title        = {AlphaInterp: Probing AlphaFold 3{\textquoteright}s Internal Representations Reveals Evolutionary Determinants of Predicted Structure and Confidence},
	author       = {Feldman, Jonathan and Skolnick, Jeffrey},
	year         = 2026,
	journal      = {bioRxiv},
	publisher    = {Cold Spring Harbor Laboratory},
	doi          = {10.64898/2026.04.22.720175},
	url          = {https://www.biorxiv.org/content/early/2026/04/23/2026.04.22.720175},
	elocation-id = {2026.04.22.720175},
	eprint       = {https://www.biorxiv.org/content/early/2026/04/23/2026.04.22.720175.full.pdf}
}

@article{Dauparas2022-su,
	title        = {Robust deep learning-based protein sequence design using {ProteinMPNN}},
	author       = {Dauparas, J and Anishchenko, I and Bennett, N and Bai, H and Ragotte, R J and Milles, L F and Wicky, B I M and Courbet, A and de Haas, R J and Bethel, N and Leung, P J Y and Huddy, T F and Pellock, S and Tischer, D and Chan, F and Koepnick, B and Nguyen, H and Kang, A and Sankaran, B and Bera, A K and King, N P and Baker, D},
	year         = 2022,
	month        = oct,
	journal      = {Science},
	publisher    = {American Association for the Advancement of Science (AAAS)},
	volume       = 378,
	number       = 6615,
	pages        = {49--56},
	language     = {en}
}

@article{Sims2005-fq,
	title        = {Protein conformational space in higher order phi-Psi maps},
	author       = {Sims, Gregory E and Choi, In-Geol and Kim, Sung-Hou},
	year         = 2005,
	month        = jan,
	journal      = {Proc. Natl. Acad. Sci. U. S. A.},
	publisher    = {Proceedings of the National Academy of Sciences},
	volume       = 102,
	number       = 3,
	pages        = {618--621},
	language     = {en}
}

@article{Pele2011-ag,
	title        = {Multidimensional scaling reveals the main evolutionary pathways of class A G-protein-coupled receptors},
	author       = {Pel{\'e}, Julien and Abdi, Herv{\'e} and Moreau, Matthieu and Thybert, David and Chabbert, Marie},
	year         = 2011,
	month        = apr,
	journal      = {PLoS One},
	publisher    = {Public Library of Science (PLoS)},
	volume       = 6,
	number       = 4,
	pages        = {e19094},
	language     = {en}
}

@article{Rotkiewicz2008-tp,
	title        = {Fast procedure for reconstruction of full-atom protein models from reduced representations},
	author       = {Rotkiewicz, Piotr and Skolnick, Jeffrey},
	year         = 2008,
	month        = jul,
	journal      = {J. Comput. Chem.},
	publisher    = {Wiley},
	volume       = 29,
	number       = 9,
	pages        = {1460--1465},
	copyright    = {http://onlinelibrary.wiley.com/termsAndConditions\#vor},
	language     = {en}
}

@article{Feldman2026-xy,
	title        = {Know your scientist: {KYC} as biosecurity infrastructure},
	author       = {Feldman, Jonathan and Feldman, Tal and Anton, Annie I},
	year         = 2026,
	month        = apr,
	journal      = {Front. Microbiol.},
	volume       = 17,
	pages        = 1814993,
	language     = {en}
}

@article{BETANCOURT2004635,
	title        = {Local Propensities and Statistical Potentials of Backbone Dihedral Angles in Proteins},
	author       = {Marcos R. Betancourt and Jeffrey Skolnick},
	year         = 2004,
	journal      = {Journal of Molecular Biology},
	volume       = 342,
	number       = 2,
	pages        = {635--649},
	doi          = {https://doi.org/10.1016/j.jmb.2004.06.091},
	issn         = {0022-2836},
	url          = {https://www.sciencedirect.com/science/article/pii/S0022283604008629}
}

@article{10.1093/bioinformatics/10.6.587,
	title        = {Flexible algorithm for direct multiple alignment of protein structures and sequences},
	author       = {Godzik, Adam and Skolnick, Jeffrey},
	year         = 1994,
	month        = 12,
	journal      = {Bioinformatics},
	volume       = 10,
	number       = 6,
	pages        = {587--596},
	doi          = {10.1093/bioinformatics/10.6.587},
	issn         = {1367-4803},
	url          = {https://doi.org/10.1093/bioinformatics/10.6.587},
	eprint       = {https://academic.oup.com/bioinformatics/article-pdf/10/6/587/812224/10-6-587.pdf}
}

@article{Kalakoti2025-rf,
	title        = {{AFsample2} predicts multiple conformations and ensembles with {AlphaFold2}},
	author       = {Kalakoti, Yogesh and Wallner, Bj{\"o}rn},
	year         = 2025,
	month        = mar,
	journal      = {Commun. Biol.},
	volume       = 8,
	number       = 1,
	pages        = 373,
	language     = {en}
}

@misc{cao2026sfclusterfrustrationguidedmsasubsampling,
	title        = {SF-Cluster: Frustration-Guided MSA Subsampling for Alternative Protein Conformation Recovery},
	author       = {Hanqun Cao and Zijun Gao and Chunbin Gu and Ge Liu and Pheng Ann Heng and Pranam Chatterjee},
	year         = 2026,
	url          = {https://arxiv.org/abs/2607.00180},
	eprint       = {2607.00180},
	archiveprefix = {arXiv},
	primaryclass = {q-bio.BM}
}

@article{Candido2026.06.03.729735,
	title        = {Language Modeling Materializes a World Model of Protein Biology},
	author       = {Candido, Salvatore and Hayes, Thomas and Derry, Alexander and Rao, Roshan and Lin, Zeming and Verkuil, Robert and Wu, Bryan Z. and Lee, Jin Sub and Bruguera, Elise S. and Keval, Jehan A. and Kopylov, Mykhailo and Pak, John E. and Wu, Wesley and Thomas, Neil and Mataraso, Samson and Hsu, Alvin and Trotman-Grant, Ashton C. and Fatras, Kilian and dos Santos Costa, Allan and Badkundri, Rohil and Ak{\i}n, Halil and Oktay, Deniz and Deaton, Jonathan and Montabana, Elizabeth and Sitwala, Hrishita and Yu, Yue and Wiggert, Marius and Carlin, Dylan Alexander and Goering, Anthony W. and Blazejewski, Tomasz and Sandora, McCullen and Hla, Michael and Jia, Tina Z. and Kloker, Leon H. and Sofroniew, Nicholas J. and Uehara, Masatoshi and Pannu, Jassi and Bachas, Sharrol and Liu, Daniel S. and Sercu, Tom and Rives, Alexander},
	year         = 2026,
	journal      = {bioRxiv},
	publisher    = {Cold Spring Harbor Laboratory},
	doi          = {10.64898/2026.06.03.729735},
	url          = {https://www.biorxiv.org/content/early/2026/06/04/2026.06.03.729735},
	elocation-id = {2026.06.03.729735},
	eprint       = {https://www.biorxiv.org/content/early/2026/06/04/2026.06.03.729735.full.pdf}
}

@misc{project2026foldingreasoningscalingopensource,
	title        = {Folding, Reasoning, and Scaling with Open-source Drug Discovery Engine},
	author       = {Aureka AI OpenDDE project},
	year         = 2026,
	url          = {https://arxiv.org/abs/2607.03787},
	eprint       = {2607.03787},
	archiveprefix = {arXiv},
	primaryclass = {cs.AI}
}

@article{Fawcett2006-wf,
	title        = {An introduction to {ROC} analysis},
	author       = {Fawcett, Tom},
	year         = 2006,
	month        = jun,
	journal      = {Pattern Recognit. Lett.},
	publisher    = {Elsevier BV},
	volume       = 27,
	number       = 8,
	pages        = {861--874},
	language     = {en}
}

@article{Wilcoxon1945-jl,
	title        = {Individual comparisons by ranking methods},
	author       = {Wilcoxon, Frank},
	year         = 1945,
	month        = dec,
	journal      = {Biom. Bull.},
	publisher    = {JSTOR},
	volume       = 1,
	number       = 6,
	pages        = 80
}

@article{Li2011-wj,
	title        = {Predicting residue-residue contacts using random forest models},
	author       = {Li, Yunqi and Fang, Yaping and Fang, Jianwen},
	year         = 2011,
	month        = dec,
	journal      = {Bioinformatics},
	publisher    = {Oxford University Press (OUP)},
	volume       = 27,
	number       = 24,
	pages        = {3379--3384},
	language     = {en}
}

@article{https://doi.org/10.1002/bip.360221211,
	title        = {Dictionary of protein secondary structure: Pattern recognition of hydrogen-bonded and geometrical features},
	author       = {Kabsch, Wolfgang and Sander, Christian},
	year         = 1983,
	journal      = {Biopolymers},
	volume       = 22,
	number       = 12,
	pages        = {2577--2637},
	doi          = {https://doi.org/10.1002/bip.360221211},
	url          = {https://onlinelibrary.wiley.com/doi/abs/10.1002/bip.360221211},
	eprint       = {https://onlinelibrary.wiley.com/doi/pdf/10.1002/bip.360221211}
}

@article{Siepen2003-ai,
	title        = {Beta edge strands in protein structure prediction and aggregation},
	author       = {Siepen, Jennifer A and Radford, Sheena E and Westhead, David R},
	year         = 2003,
	month        = oct,
	journal      = {Protein Sci.},
	publisher    = {Wiley},
	volume       = 12,
	number       = 10,
	pages        = {2348--2359},
	copyright    = {http://onlinelibrary.wiley.com/termsAndConditions\#vor},
	language     = {en}
}

@article{pacesa_one-shot_2025,
	title        = {One-shot design of functional protein binders with {BindCraft}},
	author       = {Pacesa, Martin and Nickel, Lennart and Schellhaas, Christian and Schmidt, Joseph and Pyatova, Ekaterina and Kissling, Lucas and Barendse, Patrick and Choudhury, Jagrity and Kapoor, Srajan and Alcaraz-Serna, Ana and Cho, Yehlin and Ghamary, Kourosh H. and Vinué, Laura and Yachnin, Brahm J. and Wollacott, Andrew M. and Buckley, Stephen and Westphal, Adrie H. and Lindhoud, Simon and Georgeon, Sandrine and Goverde, Casper A. and Hatzopoulos, Georgios N. and Gönczy, Pierre and Muller, Yannick D. and Schwank, Gerald and Swarts, Daan C. and Vecchio, Alex J. and Schneider, Bernard L. and Ovchinnikov, Sergey and Correia, Bruno E.},
	year         = 2025,
	month        = oct,
	journal      = {Nature},
	publisher    = {Nature Publishing Group},
	volume       = 646,
	number       = 8084,
	pages        = {483--492},
	doi          = {10.1038/s41586-025-09429-6},
	issn         = {1476-4687},
	url          = {https://www.nature.com/articles/s41586-025-09429-6},
	urldate      = {2025-12-16},
	copyright    = {2025 The Author(s)},
	language     = {en}
}

@article{hoerl_ridge_nodate,
	title        = {Ridge {Regression}: {Biased} {Estimation} for {Nonorthogonal} {Problems}},
	author       = {Hoerl, Arthur E and Kennard, Robert W},
	language     = {en}
}

@article{roy_effective_2007,
	title        = {The {Effective} {Rank}: a {Measure} of {Effective} {Dimensionality}},
	author       = {Roy, Olivier and Vetterli, Martin},
	year         = 2007,
	language     = {en}
}

@article{abramson_accurate_2024,
	title        = {Accurate structure prediction of biomolecular interactions with {AlphaFold} 3},
	author       = {Abramson, Josh and Adler, Jonas and Dunger, Jack and Evans, Richard and Green, Tim and Pritzel, Alexander and Ronneberger, Olaf and Willmore, Lindsay and Ballard, Andrew J. and Bambrick, Joshua and Bodenstein, Sebastian W. and Evans, David A. and Hung, Chia-Chun and O’Neill, Michael and Reiman, David and Tunyasuvunakool, Kathryn and Wu, Zachary and Žemgulytė, Akvilė and Arvaniti, Eirini and Beattie, Charles and Bertolli, Ottavia and Bridgland, Alex and Cherepanov, Alexey and Congreve, Miles and Cowen-Rivers, Alexander I. and Cowie, Andrew and Figurnov, Michael and Fuchs, Fabian B. and Gladman, Hannah and Jain, Rishub and Khan, Yousuf A. and Low, Caroline M. R. and Perlin, Kuba and Potapenko, Anna and Savy, Pascal and Singh, Sukhdeep and Stecula, Adrian and Thillaisundaram, Ashok and Tong, Catherine and Yakneen, Sergei and Zhong, Ellen D. and Zielinski, Michal and Žídek, Augustin and Bapst, Victor and Kohli, Pushmeet and Jaderberg, Max and Hassabis, Demis and Jumper, John M.},
	year         = 2024,
	month        = jun,
	journal      = {Nature},
	publisher    = {Nature Publishing Group},
	volume       = 630,
	number       = 8016,
	pages        = {493--500},
	doi          = {10.1038/s41586-024-07487-w},
	issn         = {1476-4687},
	url          = {https://www.nature.com/articles/s41586-024-07487-w},
	urldate      = {2025-07-21},
	copyright    = {2024 The Author(s)},
	language     = {en}
}

@article{dianzhuo_wang_without_2025,
	title        = {Without {Safeguards}, {AI}-{Biology} {Integration} {Risks} {Accelerating} {Future} {Pandemics}},
	author       = {{Dianzhuo Wang} and Huot, Marian and {Zechen Zhang} and {Kaiyi Jiang} and Shakhnovich, Eugene I and Esvelt, Kevin M},
	year         = 2025,
	publisher    = {Unpublished},
	doi          = {10.13140/RG.2.2.29765.15849},
	url          = {https://rgdoi.net/10.13140/RG.2.2.29765.15849},
	urldate      = {2025-06-22},
	language     = {en}
}

@article{biom14121531,
	title        = {The Historical Evolution and Significance of Multiple Sequence Alignment in Molecular Structure and Function Prediction},
	author       = {Zhang, Chenyue and Wang, Qinxin and Li, Yiyang and Teng, Anqi and Hu, Gang and Wuyun, Qiqige and Zheng, Wei},
	year         = 2024,
	journal      = {Biomolecules},
	volume       = 14,
	number       = 12,
	doi          = {10.3390/biom14121531},
	issn         = {2218-273X},
	url          = {https://www.mdpi.com/2218-273X/14/12/1531},
	article-number = 1531,
	pubmedid     = 39766238
}

@article{Hu2025.09.23.678064,
	title        = {OT-RMSD: A Generalization of the Root-Mean-Square Deviation for Aligning Unequal-Length Protein Structures},
	author       = {Hu, Yue and Cao, Zanxia and Liu, Yingchao},
	year         = 2025,
	journal      = {bioRxiv},
	publisher    = {Cold Spring Harbor Laboratory},
	doi          = {10.1101/2025.09.23.678064},
	url          = {https://www.biorxiv.org/content/early/2025/09/25/2025.09.23.678064},
	elocation-id = {2025.09.23.678064},
	eprint       = {https://www.biorxiv.org/content/early/2025/09/25/2025.09.23.678064.full.pdf}
}

@article{Zhao2025.05.07.652715,
	title        = {Combining structural modeling and deep learning to calculate the E. coli protein interactome and functional networks},
	author       = {Zhao, H. and Velez, C. and Navarene, A. and Saha, A. and Feldman, J. and Skolnick, J. and Murray, D. and Honig, B.},
	year         = 2025,
	journal      = {bioRxiv},
	publisher    = {Cold Spring Harbor Laboratory},
	doi          = {10.1101/2025.05.07.652715},
	url          = {https://www.biorxiv.org/content/early/2025/05/12/2025.05.07.652715},
	elocation-id = {2025.05.07.652715},
	eprint       = {https://www.biorxiv.org/content/early/2025/05/12/2025.05.07.652715.full.pdf}
}

@article{feldman_af3complex_2025,
	title        = {{AF3Complex} yields improved structural predictions of protein complexes},
	author       = {Feldman, Jonathan and Skolnick, Jeffrey},
	year         = 2025,
	month        = aug,
	journal      = {Bioinformatics},
	volume       = 41,
	number       = 8,
	pages        = {btaf432},
	doi          = {10.1093/bioinformatics/btaf432},
	issn         = {1367-4811},
	url          = {https://doi.org/10.1093/bioinformatics/btaf432},
	urldate      = {2025-10-07}
}

@misc{feldman_resilient_2025,
	title        = {Resilient {Biosecurity} in the {Era} of {AI}-{Enabled} {Bioweapons}},
	author       = {Feldman, Jonathan and Feldman, Tal},
	year         = 2025,
	month        = aug,
	publisher    = {arXiv},
	doi          = {10.48550/arXiv.2509.02610},
	url          = {http://arxiv.org/abs/2509.02610},
	urldate      = {2025-10-07},
	note         = {arXiv:2509.02610 [q-bio]}
}

@article{biasini_openstructure_2013,
	title        = {{OpenStructure}: an integrated software framework for computational structural biology},
	shorttitle   = {{OpenStructure}},
	author       = {Biasini, M. and Schmidt, T. and Bienert, S. and Mariani, V. and Studer, G. and Haas, J. and Johner, N. and Schenk, A. D. and Philippsen, A. and Schwede, T.},
	year         = 2013,
	month        = may,
	journal      = {Acta Crystallographica Section D: Biological Crystallography},
	publisher    = {International Union of Crystallography},
	volume       = 69,
	number       = 5,
	pages        = {701--709},
	doi          = {10.1107/S0907444913007051},
	issn         = {0907-4449},
	url          = {//journals.iucr.org/paper?ic5090},
	urldate      = {2024-12-03},
	language     = {en}
}

@article{tunyasuvunakool_highly_2021,
	title        = {Highly accurate protein structure prediction for the human proteome},
	author       = {Tunyasuvunakool, Kathryn and Adler, Jonas and Wu, Zachary and Green, Tim and Zielinski, Michal and Žídek, Augustin and Bridgland, Alex and Cowie, Andrew and Meyer, Clemens and Laydon, Agata and Velankar, Sameer and Kleywegt, Gerard J. and Bateman, Alex and Evans, Richard and Pritzel, Alexander and Figurnov, Michael and Ronneberger, Olaf and Bates, Russ and Kohl, Simon A. A. and Potapenko, Anna and Ballard, Andrew J. and Romera-Paredes, Bernardino and Nikolov, Stanislav and Jain, Rishub and Clancy, Ellen and Reiman, David and Petersen, Stig and Senior, Andrew W. and Kavukcuoglu, Koray and Birney, Ewan and Kohli, Pushmeet and Jumper, John and Hassabis, Demis},
	year         = 2021,
	month        = aug,
	journal      = {Nature},
	volume       = 596,
	number       = 7873,
	pages        = {590--596},
	doi          = {10.1038/s41586-021-03828-1},
	issn         = {1476-4687},
	language     = {eng}
}

@article{yang_alphafold2_2023,
	title        = {{AlphaFold2} and its applications in the fields of biology and medicine},
	author       = {Yang, Zhenyu and Zeng, Xiaoxi and Zhao, Yi and Chen, Runsheng},
	year         = 2023,
	month        = mar,
	journal      = {Signal Transduction and Targeted Therapy},
	publisher    = {Nature Publishing Group},
	volume       = 8,
	number       = 1,
	pages        = {1--14},
	doi          = {10.1038/s41392-023-01381-z},
	issn         = {2059-3635},
	url          = {https://www.nature.com/articles/s41392-023-01381-z},
	urldate      = {2024-12-03},
	copyright    = {2023 The Author(s)},
	language     = {en}
}

@article{zhang_tm-align_2005,
	title        = {{TM}-align: a protein structure alignment algorithm based on the {TM}-score},
	shorttitle   = {{TM}-align},
	author       = {Zhang, Yang and Skolnick, Jeffrey},
	year         = 2005,
	journal      = {Nucleic Acids Research},
	volume       = 33,
	number       = 7,
	pages        = {2302--2309},
	doi          = {10.1093/nar/gki524},
	issn         = {1362-4962},
	language     = {eng}
}

@article{hekkelman_alphafill_2023,
	title        = {{AlphaFill}: enriching {AlphaFold} models with ligands and cofactors},
	shorttitle   = {{AlphaFill}},
	author       = {Hekkelman, Maarten L. and de Vries, Ida and Joosten, Robbie P. and Perrakis, Anastassis},
	year         = 2023,
	month        = feb,
	journal      = {Nature Methods},
	publisher    = {Nature Publishing Group},
	volume       = 20,
	number       = 2,
	pages        = {205--213},
	doi          = {10.1038/s41592-022-01685-y},
	issn         = {1548-7105},
	url          = {https://www.nature.com/articles/s41592-022-01685-y},
	urldate      = {2024-12-02},
	copyright    = {2022 The Author(s)},
	language     = {en}
}

@article{gao_af2complex_2022,
	title        = {{AF2Complex} predicts direct physical interactions in multimeric proteins with deep learning},
	author       = {Gao, Mu and Nakajima An, Davi and Parks, Jerry M. and Skolnick, Jeffrey},
	year         = 2022,
	month        = apr,
	journal      = {Nature Communications},
	volume       = 13,
	number       = 1,
	pages        = 1744,
	doi          = {10.1038/s41467-022-29394-2},
	issn         = {2041-1723},
	url          = {https://www.nature.com/articles/s41467-022-29394-2},
	urldate      = {2024-09-14},
	language     = {en}
}

@article{roy_alphafold3_2024,
	title        = {{AlphaFold3} takes a step toward decoding molecular behavior and biological computation},
	author       = {Roy, Rohit and Al-Hashimi, Hashim M.},
	year         = 2024,
	month        = jul,
	journal      = {Nature Structural \& Molecular Biology},
	publisher    = {Nature Publishing Group},
	volume       = 31,
	number       = 7,
	pages        = {997--1000},
	doi          = {10.1038/s41594-024-01350-2},
	issn         = {1545-9985},
	url          = {https://www.nature.com/articles/s41594-024-01350-2},
	urldate      = {2024-09-01},
	copyright    = {2024 Springer Nature America, Inc.},
	language     = {en}
}

@article{skolnick_alphafold_2021,
	title        = {{AlphaFold} 2: {Why} {It} {Works} and {Its} {Implications} for {Understanding} the {Relationships} of {Protein} {Sequence}, {Structure}, and {Function}},
	shorttitle   = {{AlphaFold} 2},
	author       = {Skolnick, Jeffrey and Gao, Mu and Zhou, Hongyi and Singh, Suresh},
	year         = 2021,
	month        = oct,
	journal      = {Journal of Chemical Information and Modeling},
	volume       = 61,
	number       = 10,
	pages        = {4827--4831},
	doi          = {10.1021/acs.jcim.1c01114},
	issn         = {1549-960X},
	language     = {eng}
}

@article{jumper_highly_2021,
	title        = {Highly accurate protein structure prediction with {AlphaFold}},
	author       = {Jumper, John and Evans, Richard and Pritzel, Alexander and Green, Tim and Figurnov, Michael and Ronneberger, Olaf and Tunyasuvunakool, Kathryn and Bates, Russ and Žídek, Augustin and Potapenko, Anna and Bridgland, Alex and Meyer, Clemens and Kohl, Simon A. A. and Ballard, Andrew J. and Cowie, Andrew and Romera-Paredes, Bernardino and Nikolov, Stanislav and Jain, Rishub and Adler, Jonas and Back, Trevor and Petersen, Stig and Reiman, David and Clancy, Ellen and Zielinski, Michal and Steinegger, Martin and Pacholska, Michalina and Berghammer, Tamas and Bodenstein, Sebastian and Silver, David and Vinyals, Oriol and Senior, Andrew W. and Kavukcuoglu, Koray and Kohli, Pushmeet and Hassabis, Demis},
	year         = 2021,
	month        = aug,
	journal      = {Nature},
	volume       = 596,
	number       = 7873,
	pages        = {583--589},
	doi          = {10.1038/s41586-021-03819-2},
	issn         = {0028-0836, 1476-4687},
	url          = {https://www.nature.com/articles/s41586-021-03819-2},
	urldate      = {2024-08-31},
	language     = {en}
}

@article{10.1093/bioinformatics/bth493,
	title        = {SABmark—a benchmark for sequence alignment that covers the entire known fold space},
	author       = {Van Walle, Ivo and Lasters, Ignace and Wyns, Lode},
	year         = 2004,
	month        = {08},
	journal      = {Bioinformatics},
	volume       = 21,
	number       = 7,
	pages        = {1267--1268},
	doi          = {10.1093/bioinformatics/bth493},
	issn         = {1367-4803},
	url          = {https://doi.org/10.1093/bioinformatics/bth493},
	eprint       = {https://academic.oup.com/bioinformatics/article-pdf/21/7/1267/48966932/bioinformatics_21_7_1267.pdf}
}

@article{thadani_learning_2023,
	title        = {Learning from prepandemic data to forecast viral escape},
	author       = {Thadani, Nicole N. and Gurev, Sarah and Notin, Pascal and Youssef, Noor and Rollins, Nathan J. and Ritter, Daniel and Sander, Chris and Gal, Yarin and Marks, Debora S.},
	year         = 2023,
	month        = oct,
	journal      = {Nature},
	publisher    = {Nature Publishing Group},
	volume       = 622,
	number       = 7984,
	pages        = {818--825},
	doi          = {10.1038/s41586-023-06617-0},
	issn         = {1476-4687},
	url          = {https://www.nature.com/articles/s41586-023-06617-0},
	urldate      = {2025-05-05},
	copyright    = {2023 The Author(s)},
	language     = {en}
}

@article{feldman_biophysically_2025,
	title        = {Biophysically {Grounded} {Deep} {Learning} {Improves} {Protein}–{Protein} dd{G} {Prediction}},
	author       = {Feldman, Jonathan and Maechler, Antoine and Wang, Dianzhuo and Shakhnovich, Eugene},
	year         = 2025,
	journal      = {bioRxiv},
	publisher    = {Cold Spring Harbor Laboratory},
	doi          = {10.64898/2025.12.23.696257},
	url          = {https://www.biorxiv.org/content/early/2025/12/25/2025.12.23.696257},
	note         = {\_eprint: https://www.biorxiv.org/content/early/2025/12/25/2025.12.23.696257.full.pdf}
}

@misc{stark_boltzgen_2025,
	title        = {{BoltzGen}: {Toward} {Universal} {Binder} {Design}},
	shorttitle   = {{BoltzGen}},
	author       = {Stark, Hannes and Faltings, Felix and Choi, MinGyu and Xie, Yuxin and Hur, Eunsu and O’Donnell, Timothy and Bushuiev, Anton and Uçar, Talip and Passaro, Saro and Mao, Weian and Reveiz, Mateo and Bushuiev, Roman and Pluskal, Tomáš and Sivic, Josef and Kreis, Karsten and Vahdat, Arash and Ray, Shamayeeta and Goldstein, Jonathan T. and Savinov, Andrew and Hambalek, Jacob A. and Gupta, Anshika and Taquiri-Diaz, Diego A. and Zhang, Yaotian and Hatstat, A. Katherine and Arada, Angelika and Kim, Nam Hyeong and Tackie-Yarboi, Ethel and Boselli, Dylan and Schnaider, Lee and Liu, Chang C. and Li, Gene-Wei and Hnisz, Denes and Sabatini, David M. and DeGrado, William F. and Wohlwend, Jeremy and Corso, Gabriele and Barzilay, Regina and Jaakkola, Tommi},
	year         = 2025,
	month        = nov,
	publisher    = {bioRxiv},
	doi          = {10.1101/2025.11.20.689494},
	url          = {https://www.biorxiv.org/content/10.1101/2025.11.20.689494v1},
	urldate      = {2026-03-29},
	copyright    = {© 2025, Posted by openRxiv. This pre-print is available under a Creative Commons License (Attribution 4.0 International), CC BY 4.0, as described at http://creativecommons.org/licenses/by/4.0/},
	note         = {ISSN: 2692-8205 Pages: 2025.11.20.689494 Section: New Results},
	language     = {en}
}

@misc{cho_protein_2025,
	title        = {Protein {Hunter}: exploiting structure hallucination within diffusion for protein design},
	shorttitle   = {Protein {Hunter}},
	author       = {Cho, Yehlin and Rangel, Griffin and Bhardwaj, Gaurav and Ovchinnikov, Sergey},
	year         = 2025,
	month        = oct,
	publisher    = {bioRxiv},
	doi          = {10.1101/2025.10.10.681530},
	url          = {https://www.biorxiv.org/content/10.1101/2025.10.10.681530v1},
	urldate      = {2026-03-29},
	copyright    = {© 2025, Posted by Cold Spring Harbor Laboratory. This pre-print is available under a Creative Commons License (Attribution 4.0 International), CC BY 4.0, as described at http://creativecommons.org/licenses/by/4.0/},
	note         = {ISSN: 2692-8205 Pages: 2025.10.10.681530 Section: New Results},
	language     = {en}
}

@misc{feldman_adversarial_2026,
	title        = {Adversarial {Sequence} {Mutations} in {AlphaFold} and {ESMFold} {Reveal} {Nonphysical} {Structural} {Invariance}, {Confidence} {Failures}, and {Concerns} for {Protein} {Design}},
	author       = {Feldman, Jonathan and Brogi, Maximilian and Skolnick, Jeffrey},
	year         = 2026,
	month        = feb,
	publisher    = {bioRxiv},
	doi          = {10.64898/2026.02.25.708002},
	url          = {https://www.biorxiv.org/content/10.64898/2026.02.25.708002v1},
	urldate      = {2026-04-10},
	copyright    = {© 2026, Posted by openRxiv. This pre-print is available under a Creative Commons License (Attribution-NoDerivs 4.0 International), CC BY-ND 4.0, as described at http://creativecommons.org/licenses/by-nd/4.0/},
	note         = {ISSN: 2692-8205 Pages: 2026.02.25.708002 Section: New Results},
	language     = {en}
}

@article{https://doi.org/10.1002/pro.70208,
	title        = {DSSP 4: FAIR annotation of protein secondary structure},
	author       = {Hekkelman, Maarten L. and Salmoral, Daniel Álvarez and Perrakis, Anastassis and Joosten, Robbie P.},
	year         = 2025,
	journal      = {Protein Science},
	volume       = 34,
	number       = 8,
	pages        = {e70208},
	doi          = {https://doi.org/10.1002/pro.70208},
	url          = {https://onlinelibrary.wiley.com/doi/abs/10.1002/pro.70208},
	eprint       = {https://onlinelibrary.wiley.com/doi/pdf/10.1002/pro.70208}
}

@article{shannon_mathematical_nodate,
	title        = {A {Mathematical} {Theory} of {Communication}},
	author       = {Shannon, C E},
	language     = {en}
}

@article{rustamov_msa_2024,
	title        = {{MSA} clustering enhances {AF}-{Multimer}’s ability to predict conformational landscapes of protein–protein interactions},
	author       = {Rustamov, Khondamir R and Baev, Artyom Y},
	year         = 2024,
	month        = dec,
	journal      = {Bioinformatics Advances},
	volume       = 5,
	number       = 1,
	pages        = {vbae197},
	doi          = {10.1093/bioadv/vbae197},
	issn         = {2635-0041},
	url          = {https://pmc.ncbi.nlm.nih.gov/articles/PMC11671036/},
	urldate      = {2026-04-10}
}

\end{document}


\maketitle

\tableofcontents
\newpage

\section{The structural deficit without the MSA}
\label{si:deficit}

\begin{figure}[p]
\centering
\includegraphics[width=\textwidth,height=0.72\textheight,keepaspectratio]{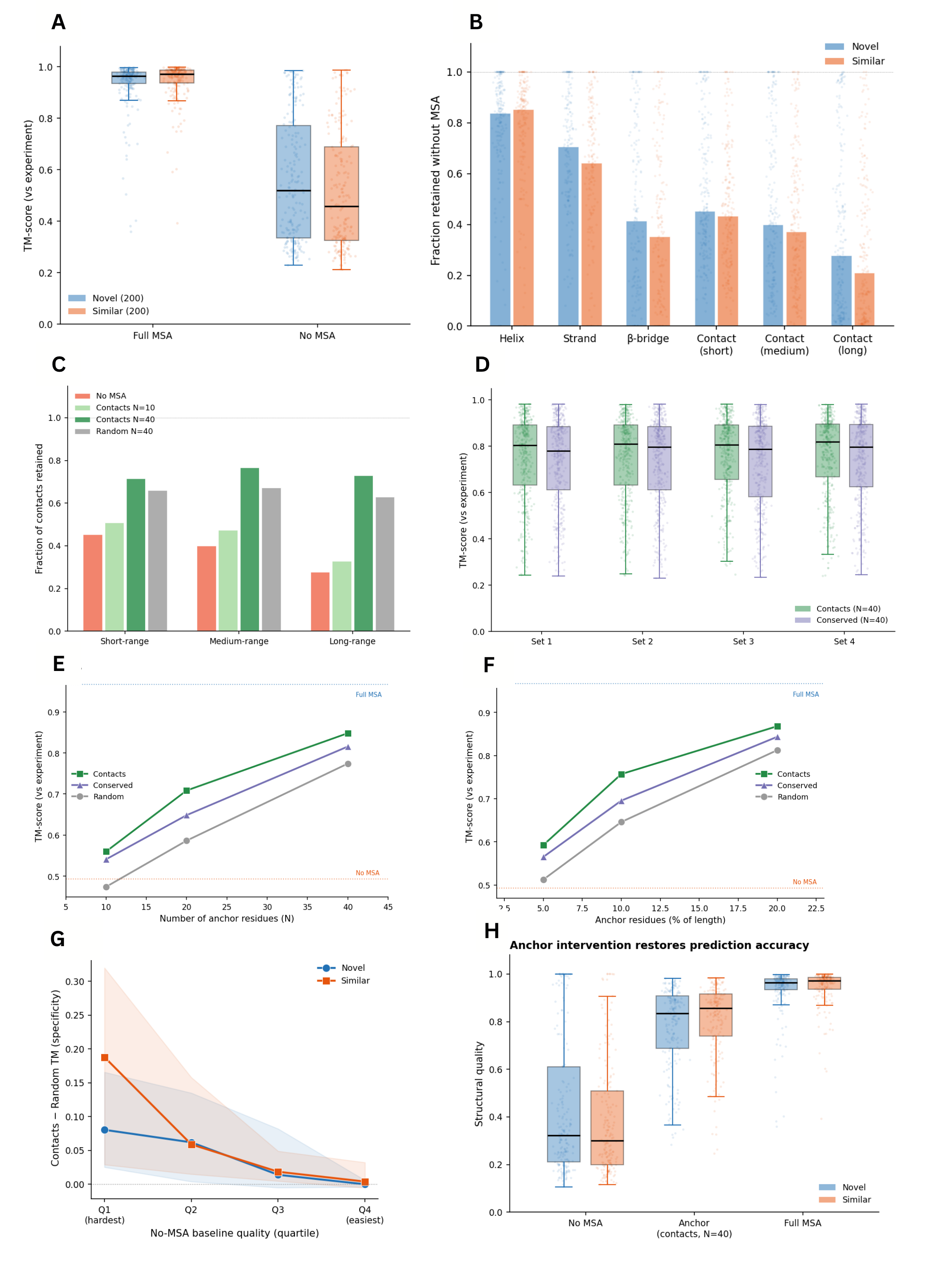}
\caption{\textbf{Evolutionary information preferentially supports nonlocal structure, which can be recovered through sparse interventions.}
\textbf{(A)} TM-score against experiment with and without the MSA in novel and similar proteins ($n=200$ each).
\textbf{(B)} Retention of helices, strands, exact $\beta$-bridge partnerships, and contacts grouped by sequence separation, measured relative to the full-MSA prediction.
\textbf{(C)} Contact retention by sequence separation in the novel cohort following MSA removal and restoration of the MSA-enriched pair representation at 10 or 40 contact-selected residues, or 40 randomly selected residues.
\textbf{(D)} TM-score against experiment following interventions at four mutually disjoint sets of 40 contact-selected or conserved residues.
\textbf{(E)} TM-score against experiment as a function of the absolute number of anchor residues---those whose MSA-enriched pair representations were restored within the no-MSA state---using contact-based, conservation-based, or random selection in both cohorts.
\textbf{(F)} Corresponding dose response when the number of anchors is specified as a percentage of sequence length. The dashed reference lines in (E) and (F) indicate full-MSA and no-MSA prediction accuracy.
\textbf{(G)} Advantage of contact-selected over randomly selected anchors in fractional TM-score recovery, stratified by the quality of the original no-MSA prediction.
\textbf{(H)} Distributions of structural quality, measured by TM-score, for the no-MSA predictions, predictions following restoration at 40 contact-selected anchors, and full-MSA predictions, shown separately for the novel and similar cohorts.
See Methods for cohort definitions, anchor selection, recovery calculations, and structural metrics.}
\label{fig:msa_restoration}
\end{figure}

Table~\ref{tab:si_retention} gives the retention measurements underlying Supplementary Figure~S1B for both cohorts. Retention is measured against each protein's own full-MSA prediction rather than against experiment, so it isolates which parts of the model's preferred structural solution survive the removal of the MSA rather than conflating that question with prediction accuracy.

\begin{table}[h]
\centering
\begin{tabular}{lcc}
\toprule
Feature & Novel & Similar \\
\midrule
Helical residues & 0.84 & 0.85 \\
Strand residues & 0.71 & 0.65 \\
Exact $\beta$-bridge partnerships & 0.42 & 0.36 \\
Contacts, short range ($6 \leq |i-j| < 12$) & 0.45 & 0.43 \\
Contacts, medium range ($12 \leq |i-j| < 24$) & 0.40 & 0.37 \\
Contacts, long range ($|i-j| \geq 24$) & 0.28 & 0.21 \\
\bottomrule \\
\end{tabular}
\caption{Fraction of each structural feature present in a protein's full-MSA prediction that is retained in its no-MSA prediction. Medians across 200 proteins per cohort.}
\label{tab:si_retention}
\end{table}

Two features of this table are worth drawing out. First, the ordering is identical in both cohorts and the gradient runs monotonically with sequence separation. Whatever the MSA supplies is therefore not distributed uniformly over the structure but concentrated in the relationships between residues distant in sequence. Second, the $\beta$-bridge measure is considerably stricter than strand retention, since it requires recovering the correct pairing partner rather than merely a strand assignment, and it is accordingly far more sensitive to MSA removal. The two measures differ by roughly thirty percentage points in both cohorts, which is to say that AlphaFold 3 without an MSA frequently produces strands it cannot pair correctly.

\section{Sparse restoration of the MSA-enriched representation}
\label{si:anchors}

\subsection{Dose response and selection strategy}

We titrated the MSA-enriched pair representation into increasing numbers of anchors, selecting them according to contact information, sequence conservation, or at random (Supplementary Figure~S1E--F). We tested absolute anchor counts of 10, 20, and 40 residues and proportional counts of 5\%, 10\%, and 20\% of sequence length under each of the three strategies. Recovery increased monotonically across the tested fixed counts and proportional doses, with 40 residues the largest fixed count examined, so the curve had not saturated at the doses we could test.

Of the 400 proteins, 328 had an experimental TM-score gap between the no-MSA and full-MSA predictions exceeding 0.1. The remaining 72 were excluded from the fractional-recovery analysis because their gaps were too small to define a stable ratio.

The advantage of contact-selected over randomly selected anchors depended strongly on how poor the no-MSA prediction was to begin with (Supplementary Figure~S1G). Stratifying by the quality of that baseline, the advantage was greatest in the hardest quartile and diminished monotonically as baseline accuracy improved, approaching zero in the easiest quartile. Where the model already recovers much of the fold without evolutionary information, it matters little which residues are enriched; where it does not, the choice matters a great deal.

Contact retention by sequence separation under each intervention is shown in Supplementary Figure~S1C. In the novel cohort, long-range-contact retention relative to the full-MSA prediction was 28\% without the MSA, 33\% with 10 contact-selected anchors, 63\% with 40 random anchors, and 73\% with 40 contact-selected anchors.

\subsection{Recovery beyond the restored pairs}

Because selecting an anchor replaces its pairwise representations with every other residue, recovery among pairs containing an anchor could arise directly from the injected information. To separate direct from indirect recovery, we divided long-range residue pairs into three categories: UU, where neither residue was selected as an anchor; IU, where exactly one was; and II, where both were (see Methods).

Approximately 68\% of all long-range pairs fell into the untouched UU category, in which no portion of the pair entry was directly replaced. Restoration of 40 contact-selected anchors nevertheless closed 34.9\% of the available gap in balanced contact-map accuracy for these pairs in the novel cohort, and 49.4\% in the similar cohort. Recovery rose to 50.6\% and 64.6\% for IU pairs, and to 62.9\% and 81.0\% for II pairs.

\begin{table}[h]
\centering
\begin{tabular}{lcc}
\toprule
Pair category & Novel & Similar \\
\midrule
UU (neither residue an anchor) & 0.349 & 0.494 \\
IU (one residue an anchor) & 0.506 & 0.646 \\
II (both residues anchors) & 0.629 & 0.810 \\
\bottomrule \\
\end{tabular}
\caption{Fraction of the gap in balanced long-range contact-map accuracy between the no-MSA and full-MSA predictions closed by restoration at 40 contact-selected anchors, by whether the pair's representation was directly modified. Medians across 200 proteins per cohort.}
\label{tab:si_uu}
\end{table}

Thus, although directly restoring the representations associated with a pair produced greater recovery, substantial recovery still occurred among pairs that were never explicitly enriched. The Pairformer and the downstream structure-generation modules can therefore use the restored information to reconstruct additional relationships throughout the protein, rather than merely transcribing what was injected.

\section{The MSA module's contribution}
\label{si:module}

\begin{figure}[p]
\centering
\includegraphics[width=\textwidth,height=0.72\textheight,keepaspectratio]{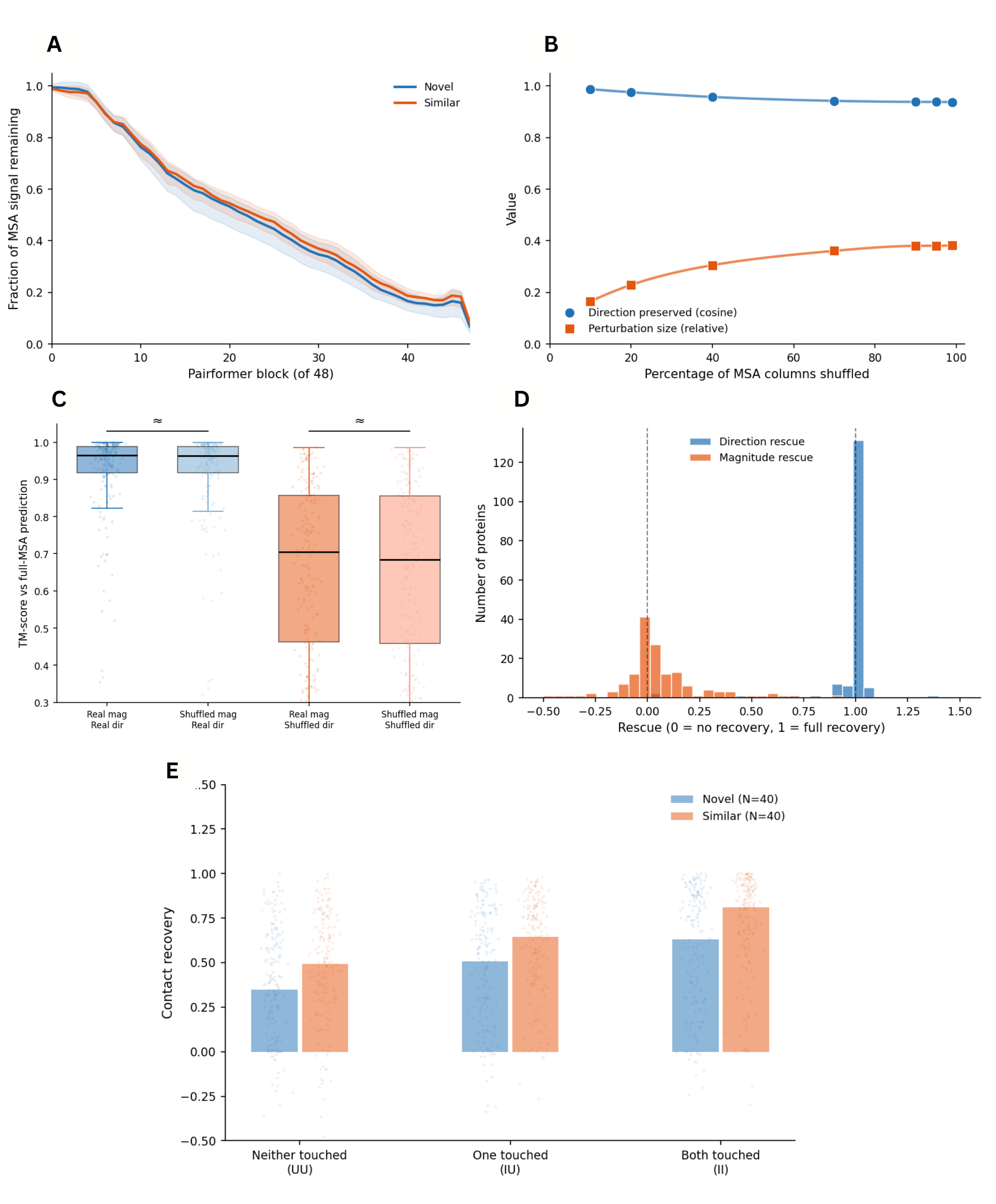}
\caption{\textbf{Cross-column evolutionary relationships modify the MSA-derived representation, whose detailed direction influences structural prediction and whose partial restoration recovers relationships beyond directly modified residue pairs.}
The MSA-specific module contribution is defined as
$M=(P_B^{\mathrm{MSA}}-P_A^{\mathrm{MSA}})-(P_B^{\mathrm{noMSA}}-P_A^{\mathrm{noMSA}})$,
distinct from the checkpoint-B difference $\Delta_B$ used for anchor interventions. The magnitude--direction swaps instead use the raw module updates from the native and shuffled conditions.
\textbf{(A)} Persistence of the original MSA-specific contribution across Pairformer blocks in the novel and similar cohorts, measured by its projection onto the evolving MSA-dependent pair representation. The same quantity is shown for the novel cohort in Figure~3A.
\textbf{(B)} Cosine similarity and relative perturbation of the MSA-derived representation following within-column MSA shuffling, which preserves individual-column statistics while disrupting cross-column evolutionary relationships.
\textbf{(C)} TM-score against the full-MSA prediction following exchange of magnitude and direction between the native and 99\%-shuffled module updates. The four conditions combine native or shuffled magnitude with native or shuffled direction.
\textbf{(D)} Per-protein direction and magnitude rescue among the 157 novel proteins meeting the minimum structural-difference criterion. Restoring the native direction recovers much of the native structural solution within the tested perturbations, whereas restoring the native magnitude alone produces little recovery.
\textbf{(E)} Fractional recovery of balanced long-range contact-map accuracy following restoration at 40 contact-selected anchors, separated into pairs with neither residue selected (UU), exactly one residue selected (IU), or both residues selected (II). This panel accompanies Section~S2.}
\label{fig:msa_module_write}
\end{figure}

\subsection{Spatial organization}

For each residue pair, the MSA module produces a 128-dimensional update, which we decomposed into its magnitude, describing the overall size of the update, and its direction, describing the relative contributions of the constituent dimensions (see Methods). The spatial organization of the magnitude was remarkably simple: a median effective rank of 1.15, with 95.1\% of the spatial spectral mass captured by the first ten components. The representation is therefore concentrated in a small number of spatial patterns rather than distributed arbitrarily across residue pairs.

Residues with greater evolutionary variability and greater alignment coverage tended to receive larger representational updates, suggesting that the magnitude of the MSA-derived representation reflects, at least in part, the amount and distribution of evolutionary evidence available at individual positions.

\subsection{Disrupting cross-column relationships}

To distinguish the contribution of cross-column evolutionary relationships from that of amino-acid composition and conservation, we independently shuffled amino acids among homologous sequences within selected MSA columns, preserving the query sequence, column positions, amino-acid frequencies, variability, and gap frequencies, while disrupting which amino acids occurred together within the same homologous sequence (see Methods).

The representation deviated further from its unshuffled counterpart as more columns were shuffled (Supplementary Figure~S2B). The overall cosine similarity between representations nevertheless remained high, at 0.938, even after shuffling 99\% of eligible columns. Much of the original representational direction was therefore preserved, and yet this apparent similarity did not extend to the resulting structures: the contact-map overlap with the unshuffled prediction fell to a Jaccard index of 0.740.

Disrupting cross-column relationships thus produced changes that were modest when assessed by the overall orientation of the representation but substantial in the contact relationships of the resulting predictions. The information written by the MSA module cannot be explained by the composition or conservation of individual columns alone.

\subsection{Magnitude and direction}

We next asked whether the structural consequences of shuffling arose from changes in the magnitude of the module update, its direction in 128-dimensional channel space, or both. We decomposed the native and shuffled updates into magnitude and direction, constructed four combinations by swapping these components between conditions, and inserted each into a no-MSA forward pass (see Methods; Supplementary Figure~S2C).

\begin{table}[h]
\centering
\begin{tabular}{llc}
\toprule
Magnitude & Direction & TM-score vs full-MSA prediction \\
\midrule
Native & Native & 0.96 \\
Shuffled & Native & 0.96 \\
Native & Shuffled & 0.70 \\
Shuffled & Shuffled & 0.68 \\
\bottomrule \\
\end{tabular}
\caption{Median TM-score against the full-MSA prediction for the four recombinations of magnitude and direction between the native and 99\%-shuffled MSA-module updates, novel cohort.}
\label{tab:si_swap}
\end{table}

The two properties had markedly different effects. Native direction with shuffled magnitude was indistinguishable from the fully native control, while native magnitude with shuffled direction was barely above the fully shuffled condition.

We also quantified, per protein, the fraction of the native-versus-shuffled structural difference recovered by each component (Supplementary Figure~S2D; see Methods). Among the 157 novel proteins for which shuffling produced a difference large enough to define a stable rescue ratio, restoring the native direction recovered a median of 1.000 of the available difference, with 154 of 157 proteins (98\%) exceeding 0.5 rescue. Restoring the native magnitude recovered a median of 0.011. The remaining 43 novel proteins failed the minimum-span criterion; corresponding swaps were not generated for the similar cohort.

Within the conditions examined, the structural outcome therefore depended predominantly on the detailed direction of the MSA-module representation rather than its overall magnitude.

\section{Structural progression of the diffusion readouts}
\label{si:protofold}

\begin{figure}[p]
\centering
\includegraphics[width=\textwidth,keepaspectratio]{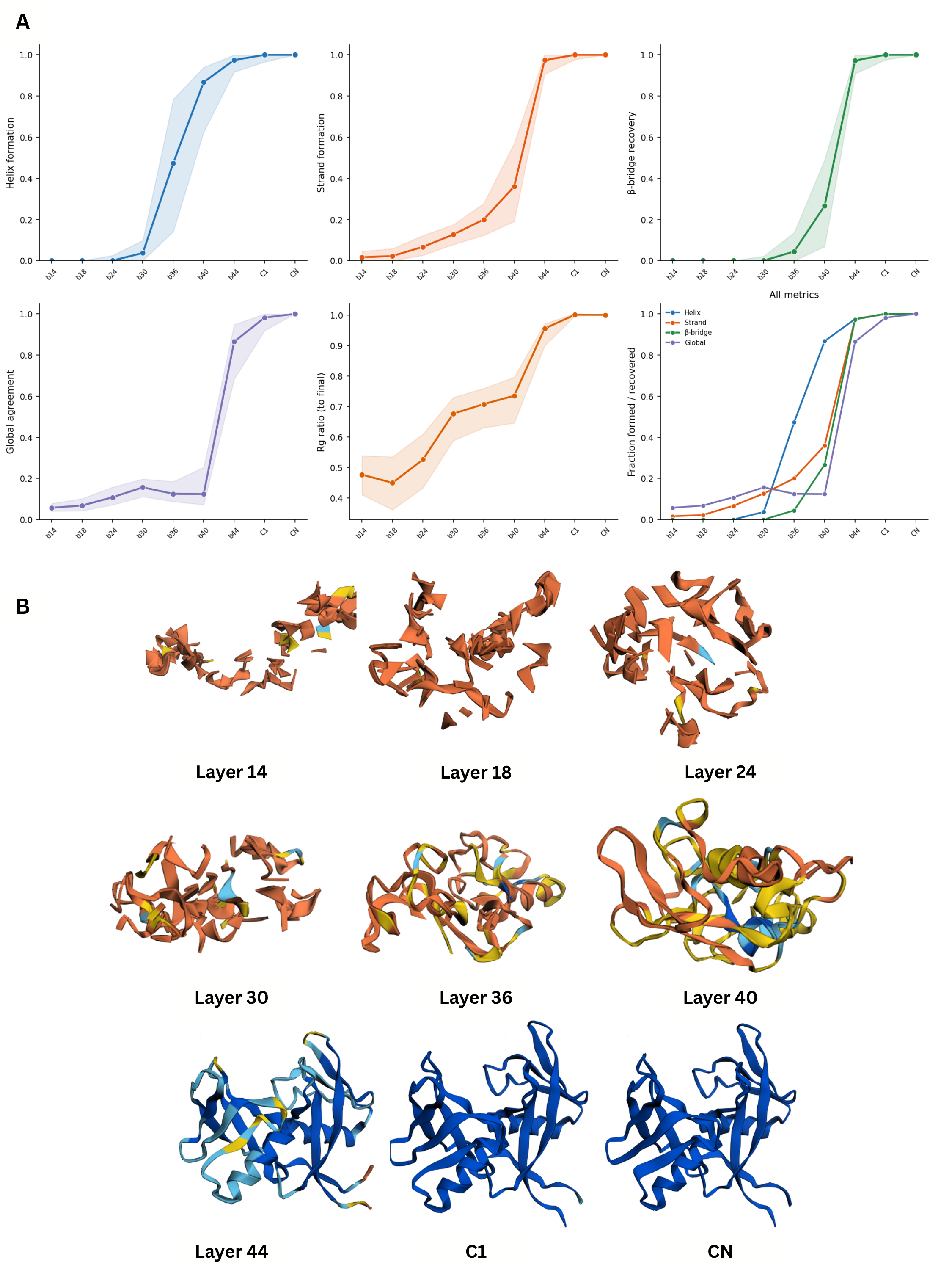}
\caption{\textbf{Structural features of the diffusion readouts across the first Pairformer pass.}
Measurements relative to the final recycled full-MSA prediction ($C_N$), across Pairformer blocks 14, 18, 24, 30, 36, 40, and 44, $C_1$, and $C_N$: helix formation, strand formation, exact $\beta$-bridge recovery, global structural agreement, and the radius of gyration as a ratio of its final value. The final panel overlays the median helix, strand, $\beta$-bridge, and global-agreement curves. Lines are medians and shaded bands are interquartile ranges.}
\label{fig:si_protofold}
\end{figure}

\subsection{The order in which features appear}

As the first Pairformer pass proceeds, the diffusion readouts expand and recover recognizable structure, but the features do not appear together. The radius-of-gyration ratio rose from 0.48 at block~14 to 0.73 at block~40. Over the same interval, helix recovery increased from 4\% at block~30 to 87\% by block~40, whereas strand recovery reached only 36\% and exact $\beta$-bridge partnerships only 27\% at that same block.

The ordering is therefore helices, then strands, then the correct pairing between strands, with global agreement resolving last. It mirrors the hierarchy of MSA dependence reported in Section~S1: the features that survive MSA removal are the ones that appear earliest here, and the features that collapse without the MSA are the ones that appear last. Both orderings separate local secondary structure from the nonlocal relationships that organize it.

\subsection{The late transition depends on the MSA}

Comparing readouts generated with and without the MSA at the same blocks shows that this progression is not a property of the diffusion module alone. With the MSA, the median TM-score against experiment rose from 0.246 at block~18 to 0.650 at block~40, then sharply to 0.923 by block~44. Without the MSA, accuracy improved far more gradually, from 0.210 at block~18 to 0.320 at block~40 and 0.354 at block~44.

\begin{table}[h]
\centering
\begin{tabular}{lcc}
\toprule
Block & With MSA & Without MSA \\
\midrule
18 & 0.246 & 0.210 \\
40 & 0.650 & 0.320 \\
44 & 0.923 & 0.354 \\
\bottomrule \\
\end{tabular}
\caption{Median TM-score against experiment for structures diffused independently from saved Pairformer representations, novel cohort. The pronounced late transition between blocks~40 and~44 occurs only when evolutionary information is available.}
\label{tab:si_diffusion}
\end{table}

The sharp improvement between blocks~40 and~44 is thus conditional on the MSA. Without it, no comparable transition occurs anywhere in the first pass.

\section{Probe controls}
\label{si:probe_controls}

Three controls establish that the decoded structures reflect residue-specific information in the pair representation rather than artifacts of the decoding procedure.

\textbf{Generic-chain floor.} We constructed a baseline containing no information about any particular fold. From the final structures of the similar cohort, we computed the mean C$\alpha$--C$\alpha$ distance at each sequence separation within five length bins, assigned each novel protein the distance profile of the nearest length bin, and reconstructed and evaluated it exactly as for the probe-decoded distances. This baseline reached a median TM-score of 0.22 against the final structures. Any reconstruction scoring at or near 0.22 therefore carries no more structural information than the generic statistics of polypeptide chains of that length.

\textbf{Residue permutation.} As a control for leakage, we randomly permuted the residues of the block-47 representation before decoding while keeping the target distances fixed. The coefficient of determination fell below zero for every combination of probe and representation, confirming that the probes read structure from the identity and arrangement of specific residue pairs rather than from generic distance statistics that would survive permutation.

\textbf{Cross-arm application.} Each probe was additionally applied to representations of the opposite type, as a control for differences in representational format between the two conditions. The probe trained without the MSA recovered a median TM-score of 0.62 at block~47 when applied to with-MSA representations, against 0.29 on its own. The features it learned are therefore decodable in principle; they are absent from the no-MSA representation.

Reconstructions are interpolations of the global arrangement implied by the decoded distances. They are not diffused, so their local geometry is imperfect and individual residues deviate from their final positions even in the best reconstructions. All reported measures were computed from the unrefined classical reconstructions.

\section{Determinants of decoding order}
\label{si:order}

We asked whether any property of the alignment or of the structure determined which residues became decodable first. For each residue we defined an earliness score from the first layer at which its long-range lDDT reached half of its final value and held, and related that score to residue burial, alignment gap fraction, column conservation, and covariation (see Methods).

\begin{table}[h]
\centering
\begin{tabular}{lc}
\toprule
Feature & Median within-protein Spearman $\rho$ \\
\midrule
Residue burial & $+0.39$ \\
Alignment gap fraction & $-0.34$ \\
Conservation & $+0.15$ \\
Covariation & $+0.13$ \\
\midrule
All features combined (cross-validated) & $0.47$ \\
\bottomrule \\
\end{tabular}
\caption{Association between per-residue features and the order in which residues become decodable. Positive values indicate earlier decoding. Burial was positively associated in every protein examined; gap fraction was negatively associated in 93\%.}
\label{tab:si_order}
\end{table}

Buried residues in well-aligned columns tended to be decoded earliest, but the associations are not strong enough to be explanatory: the two alignment statistics most often invoked to explain the MSA's contribution, conservation and covariation, were the weakest of the four, and all features combined reached only 0.47.

We further asked whether the earliest-decoded residues formed a compact spatial cluster, as a folding nucleus would. They did cluster, with a mean pairwise distance below that of random residue sets of the same size. However, early residues are also disproportionately buried, and buried residues are necessarily near one another and near the protein's centre. Against random sets matched to the same distribution of burial, the clustering was no longer significant. We therefore find no evidence for a distinct folding nucleus in AlphaFold 3's computation.

Finally, the order in which residues become decodable from the representation is unrelated to the order in which the diffusion module places them (Spearman $\rho \approx 0$). The two readouts resolve the structure in different sequences, which is further evidence that the compaction described in Section~S4 belongs to the decoder rather than to the representation.

\section{Diffusion readouts of the disjoint MSA subsets}
\label{si:disjoint_diffusion}

Table~\ref{tab:si_disjoint_diffusion} gives the diffusion readouts for the two disjoint MSA subsets, the lens superseded by the probe reconstructions in the main text.

\begin{table}[h]
\centering
\begin{tabular}{lccc}
\toprule
Block & A vs B & A vs Exp & B vs Exp \\
\midrule
14  & 0.712 & 0.197 & 0.194 \\
18  & 0.748 & 0.194 & 0.192 \\
24  & 0.723 & 0.220 & 0.224 \\
30  & 0.700 & 0.277 & 0.268 \\
36  & 0.707 & 0.407 & 0.399 \\
40  & 0.772 & 0.504 & 0.506 \\
44  & 0.808 & 0.796 & 0.768 \\
$C_N$ & 0.810 & 0.841 & 0.818 \\
\bottomrule \\
\end{tabular}
\caption{Median TM-scores for structures generated by the diffusion module from the representations of two evolutionarily independent MSA subsets, compared between subsets and against the experimental structure, for 100 novel proteins.}
\label{tab:si_disjoint_diffusion}
\end{table}

The two subsets agree at a TM-score of 0.71 as early as block~14, and that agreement barely changes until block~40. Taken at face value this would suggest the subsets follow nearly identical trajectories. The agreement is an artifact of the readout: through block~36 both predictions are overcompact and poorly organized proto-folds (Section~S4), and two compact masses of comparable size resemble one another regardless of what produced them. Agreement between the readouts is therefore interpretable only from block~40 onward, where both have resolved into recognizable structures, and the final predictions differ at 0.81.

This limitation motivated the probe-based comparison in the main text, which reads the representation directly and is unaffected by the decoder's behavior on intermediate states. Under that lens the two subsets agree at only 0.58 at block~14, and the ordering with respect to query substitution reverses.

\begin{table}[p]
\centering
\small
\begin{tabular}{lccccc}
\toprule
 & \multicolumn{2}{c}{Sabmark} & \multicolumn{2}{c}{Novel cohort} & \\
\cmidrule(lr){2-3}\cmidrule(lr){4-5}
Condition & TM & gap closed & TM & gap closed & $n$ \\
\midrule
Own full MSA (ceiling)             & 0.951 &     --- & 0.916 &     --- & 94 / 38 \\
Own MSA, aligned columns           & 0.836 &   +0.75 & 0.884 &   +0.92 & 94 / 38 \\
Partner's MSA, aligned columns     & 0.717 &   +0.33 & 0.762 &   +0.43 & 94 / 38 \\
Partner's MSA, ungapped            & 0.743 &   +0.46 & 0.762 &   +0.43 & 94 / 38 \\
Partner's MSA, contiguous blocks   & 0.846 &   +0.79 &   --- &     --- & 20 / --- \\
No MSA (floor)                     & 0.442 &     --- & 0.568 &     --- & 94 / 38 \\
Own MSA, aligned columns masked    & 0.570 &   +0.15 &   --- &     --- & 94 / --- \\
Own MSA, aligned columns corrupted & 0.519 &   +0.09 & 0.479 & $-$0.19 & 94 / 38 \\
Register shifted 3                 & 0.425 & $-$0.04 & 0.538 & $-$0.21 & 94 / 38 \\
Register shifted 7                 & 0.397 & $-$0.07 &   --- &     --- & 94 / --- \\
Register scrambled                 & 0.399 & $-$0.05 & 0.456 & $-$0.26 & 94 / 38 \\
Donor complement                   & 0.407 & $-$0.02 &   --- &     --- & 94 / --- \\
\bottomrule \\
\end{tabular}
\vspace{0.5em}
\caption{\textbf{Every transplant condition and null.} Median TM-score against the
experimental structure, and the median fraction of the gap between the no-MSA and
own-full-MSA predictions that each condition closes. Conditions run on only a subset
of pairs are marked accordingly in $n$ and are not directly comparable with the
pooled values. Negative fractions indicate conditions that perform worse than
supplying no alignment at all.}
\label{tab:transplant_all}
\end{table}

\begin{table}[p]
\centering
\small
\begin{tabular}{lcccccc}
\toprule
Block & Own full & Own aligned & Partner aligned & Partner ungapped & Contiguous & No MSA \\
\midrule
0 & 0.274 & 0.218 & 0.224 & 0.233 & 0.240 & 0.203 \\
5 & 0.433 & 0.282 & 0.278 & 0.299 & 0.324 & 0.227 \\
9 & 0.557 & 0.326 & 0.319 & 0.362 & 0.432 & 0.243 \\
14 & 0.601 & 0.355 & 0.342 & 0.426 & 0.506 & 0.248 \\
24 & 0.612 & 0.389 & 0.356 & 0.428 & 0.517 & 0.248 \\
30 & 0.698 & 0.494 & 0.410 & 0.474 & 0.558 & 0.251 \\
36 & 0.682 & 0.529 & 0.443 & 0.513 & 0.584 & 0.257 \\
40 & 0.775 & 0.636 & 0.508 & 0.581 & 0.657 & 0.264 \\
44 & 0.761 & 0.662 & 0.536 & 0.573 & 0.647 & 0.265 \\
47 & 0.793 & 0.671 & 0.483 & 0.539 & 0.648 & 0.269 \\
$C_N$ & 1.000 & 0.874 & 0.721 & 0.765 & 0.884 & 0.451 \\
\bottomrule \\
\end{tabular}
\vspace{0.5em}
\caption{\textbf{Sabmark: fold decodability across the Pairformer, transplant conditions.}
Median TM-score of the probe reconstruction against each target's own final full-MSA
prediction, which is therefore 1.000 for that condition by construction. $C_N$ is the
diffusion readout of the block-47 representation. The contiguous-block transplant was
run only on the 20 higher-coverage targets.}
\label{tab:si_sab_transplant}
\end{table}

\begin{table}[p]
\centering
\small
\begin{tabular}{lcccccccc}
\toprule
Block & Own full & Shift 3 & Shift 7 & Scrambled & Masked & Corrupted & Complement & No MSA \\
\midrule
0 & 0.274 & 0.210 & 0.198 & 0.211 & 0.189 & 0.202 & 0.186 & 0.203 \\
5 & 0.433 & 0.224 & 0.210 & 0.223 & 0.197 & 0.210 & 0.208 & 0.227 \\
9 & 0.557 & 0.243 & 0.231 & 0.243 & 0.185 & 0.214 & 0.233 & 0.243 \\
14 & 0.601 & 0.232 & 0.219 & 0.236 & 0.195 & 0.220 & 0.219 & 0.248 \\
24 & 0.612 & 0.220 & 0.200 & 0.221 & 0.194 & 0.215 & 0.204 & 0.248 \\
30 & 0.698 & 0.234 & 0.220 & 0.233 & 0.249 & 0.243 & 0.218 & 0.251 \\
36 & 0.682 & 0.235 & 0.222 & 0.236 & 0.269 & 0.267 & 0.220 & 0.257 \\
40 & 0.775 & 0.261 & 0.254 & 0.274 & 0.356 & 0.313 & 0.253 & 0.264 \\
44 & 0.761 & 0.266 & 0.257 & 0.266 & 0.355 & 0.311 & 0.254 & 0.265 \\
47 & 0.793 & 0.258 & 0.249 & 0.255 & 0.340 & 0.312 & 0.249 & 0.269 \\
$C_N$ & 1.000 & 0.411 & 0.392 & 0.402 & 0.606 & 0.516 & 0.394 & 0.451 \\
\bottomrule \\
\end{tabular}
\vspace{0.5em}
\caption{\textbf{Sabmark: fold decodability across the Pairformer, nulls.} Each null
supplies the same donor alignment at the same depth and occupancy, altering only where
its columns land. Scores are against each target's own final full-MSA prediction, as in
Table~\ref{tab:si_sab_transplant}. No null produces a decodable fold in any protein
before block~27.}
\label{tab:si_sab_nulls}
\end{table}

\begin{table}[p]
\centering
\small
\begin{tabular}{lcccccccc}
\toprule
Block & Own full & Own aligned & Partner aligned & Partner ungapped & No MSA & Corrupted & Shift 3 & Scrambled \\
\midrule
0 & 0.267 & 0.243 & 0.251 & 0.253 & 0.230 & 0.231 & 0.242 & 0.232 \\
5 & 0.404 & 0.348 & 0.290 & 0.294 & 0.230 & 0.218 & 0.250 & 0.230 \\
9 & 0.683 & 0.480 & 0.424 & 0.431 & 0.245 & 0.247 & 0.270 & 0.251 \\
14 & 0.700 & 0.553 & 0.477 & 0.483 & 0.269 & 0.243 & 0.270 & 0.266 \\
24 & 0.707 & 0.578 & 0.504 & 0.497 & 0.287 & 0.255 & 0.275 & 0.267 \\
30 & 0.783 & 0.662 & 0.555 & 0.552 & 0.299 & 0.275 & 0.315 & 0.279 \\
36 & 0.809 & 0.717 & 0.593 & 0.592 & 0.302 & 0.282 & 0.348 & 0.302 \\
40 & 0.861 & 0.781 & 0.679 & 0.679 & 0.302 & 0.335 & 0.414 & 0.339 \\
44 & 0.854 & 0.801 & 0.683 & 0.683 & 0.314 & 0.330 & 0.398 & 0.351 \\
47 & 0.855 & 0.807 & 0.700 & 0.699 & 0.300 & 0.322 & 0.406 & 0.345 \\
$C_N$ & 1.000 & 0.930 & 0.792 & 0.788 & 0.574 & 0.484 & 0.556 & 0.466 \\
\bottomrule \\
\end{tabular}
\vspace{0.5em}
\caption{\textbf{Novel cohort: fold decodability across the Pairformer.} Transplant
conditions and nulls for the 38 post-cutoff targets, scored against each target's own
final full-MSA prediction. $C_N$ is the diffusion readout of the block-47
representation.}
\label{tab:si_novel}
\end{table}

\begin{table}[p]
\centering
\small
\begin{tabular}{lcccccccc}
\toprule
Coverage & $n$ & Median & Aligned & No MSA & Transplant & Own aligned & Own full & Gap \\
quartile &     & coverage & residues &        &            &             &          & closed \\
\midrule
0.26--0.51 & 24 & 0.41 & 69 & 0.439 & 0.567 & 0.721 & 0.958 & +0.05 \\
0.51--0.60 & 23 & 0.56 & 87 & 0.389 & 0.690 & 0.808 & 0.945 & +0.44 \\
0.60--0.72 & 24 & 0.67 & 80 & 0.543 & 0.735 & 0.848 & 0.940 & +0.32 \\
0.72--0.93 & 23 & 0.79 & 80 & 0.451 & 0.856 & 0.907 & 0.953 & +0.79 \\
\bottomrule \\
\end{tabular}
\vspace{0.5em}
\caption{\textbf{The transplant recovers as much of the fold as the donor's alignment
covers.} Sabmark targets binned by the fraction of the target chain covered by the
structural alignment. Median TM-scores against the experimental structure. Across
targets, the Spearman correlation between coverage and transplant accuracy is $+0.51$
($p = 1\times10^{-7}$), and between sequence identity and transplant accuracy $+0.35$
($p = 6\times10^{-4}$). Controlling for identity, coverage retains a partial correlation
of $+0.45$; controlling for coverage, identity retains $+0.25$. Coverage is therefore
the stronger determinant.}
\label{tab:transplant_coverage}
\end{table}

\section{Supplementary Methods}
\label{si:methods}

Methods for analyses reported only in the Supplementary Information are given here. Further information about methods in the manuscript are also provided here, such as detailed algorithms. All conventions, metrics, and prediction settings follow the main-text Methods unless stated otherwise.

\subsection{Anchor titration}
\label{si:titration}

For the titration analysis (Supplementary Figure~S1E--F), we varied the number of anchors, $k$, across 10, 20, and 40 residues, as well as 5\%, 10\%, and 20\% of sequence length, for each of the three selection strategies defined in the main-text Methods. Each combination was run independently for every protein in both cohorts.

Fractional recovery was computed against the experimental TM-score gap between the no-MSA and full-MSA predictions of the same protein. Proteins whose gap did not exceed 0.1 were excluded, as the ratio is unstable when the denominator approaches zero; 328 of 400 proteins met this criterion.

To examine how the advantage of informed anchor selection varied with the difficulty of the prediction, we stratified proteins into quartiles by the TM-score of their no-MSA prediction against experiment and, within each quartile, computed the difference in fractional recovery between contact-selected and randomly selected anchors at the same dose.

\subsection{Touched and untouched pair categorization}
\label{si:uu}

To assess whether structural recovery extended beyond the pair representations directly restored by the intervention, we classified all long-range residue pairs by their overlap with the anchor set, $\mathcal{A}$: II, where both residues were anchors; IU, where exactly one was; and UU, where neither was, so that the corresponding entry of the pair representation was never directly modified. Within each category, we measured agreement with the full-MSA prediction as the balanced accuracy of the long-range contact map (C$\beta$--C$\beta$ distance below 8~\AA{}, $|i-j| \geq 24$), the mean of the fractions of reference contacts and reference non-contacts reproduced, so that the large majority of non-contacts could not dominate the measure. Recovery was the fraction of the gap between the no-MSA and full-MSA predictions closed by the intervention,
\[
\mathrm{recovery} = \frac{a_{\mathrm{anchor}} - a_{\mathrm{noMSA}}}{a_{\mathrm{MSA}} - a_{\mathrm{noMSA}}},
\]
where $a$ denotes balanced agreement within that category.

The intervention itself was not restricted to long-range pairs: selecting an anchor restored its complete row and column of the pair representation, including all 128 channels and its relationships with every other residue. The II, IU, and UU categories served only to evaluate how recovery varied with whether a long-range pair had been directly modified.

\subsection{Structural element progression}
\label{si:progression}

To characterize how recognizable structural features emerged across the Pairformer, we measured five properties of each intermediate diffusion readout relative to the final recycled full-MSA prediction, $C_N$: helix retention, strand retention, exact $\beta$-bridge recovery, global structural agreement, and the radius-of-gyration ratio, each as defined in the main-text Methods. All measurements were computed separately for each protein and reported as cohort medians.

The same measurements were computed for readouts generated under the no-MSA condition at the corresponding checkpoints, excluding block~14, and compared block by block against their full-MSA counterparts for the same proteins.

\subsection{Determinants of decoding order}
\label{si:order_methods}

To identify which residues became decodable first, we defined for each residue the earliest layer at which its long-range lDDT reached half of its value at block~47 and remained at or above that level through block~47, and used the negative of this layer as an earliness score, so that higher values indicate earlier decoding. Residues whose long-range lDDT at block~47 was below 0.3 were excluded, as their timing is not meaningfully defined.

We related this earliness score to four per-residue features: residue burial, defined as the number of C$\beta$ atoms (C$\alpha$ for glycine) within 14~\AA{} in the final structure; alignment gap fraction, the proportion of alignment rows carrying a gap at that column; conservation, defined as $1 - H/\ln 20$, where $H$ is the Shannon entropy of the alignment column; and covariation, defined as the strongest average-product-corrected mutual information between that residue's column and the column of any long-range partner, computed from up to 2,000 sequences of the full alignment. For each feature we computed the Spearman correlation with earliness within each protein and report the median across proteins, so that proteins contribute equally regardless of length.

To estimate how much of the ordering these features explain jointly, we fitted ridge-regression models to within-protein ranks with five-fold cross-validation grouped by protein, and report the median per-protein Spearman correlation between predicted and observed order.

To test whether the earliest-decoded residues formed a spatially compact cluster, we compared the mean pairwise distance among the earliest 15\% of residues against two nulls: random residue sets of the same size, and random sets matched to the same distribution of burial. The second null is the informative one, since buried residues are necessarily close to one another and to the protein's centre. We additionally computed the Spearman correlation between each residue's earliness score and the order in which the diffusion module placed it, defined analogously from the sequence of diffusion readouts.

\subsection{Contact enrichment in the orthogonal residual}
\label{si:enrichment}

To ask whether the component of the pair representation orthogonal to the original MSA-module contribution carried structural information, we measured contact enrichment within it. At each Pairformer block we computed the residual $R^{(l)}$ as defined in the main-text Methods, and took the Euclidean norm of its 128-dimensional entry at each residue pair. We then compared the median norm at long-range pairs in contact in the final full-MSA prediction (C$\beta$--C$\beta$ distance below 8~\AA{}, $|i-j| \geq 24$) with the median norm at long-range pairs not in contact. Contact enrichment is the ratio of these two medians, so that a value of one indicates no preference and values above one indicate that the residual is concentrated at the pairs that end in contact. Values were computed per protein and reported as cohort medians. Enrichment rose between blocks~20 and~40, indicating that the growth of the orthogonal component is not a loss of the MSA's contribution but its re-expression in a form increasingly aligned with the predicted contact map.

\subsection{Transplant and ablation algorithms}
\label{si:transplant_algorithms}

Each algorithm below takes the structural alignment $\mathcal{A} = \{(i,j)\}$ between target $A$ and donor $B$, where $i$ indexes a column of $A$'s alignment and $j$ a column of $B$'s, together with the gap-stripped rows of $B$'s unpaired alignment. Every algorithm returns a set of rows of length $L_A$, to which $A$'s own query sequence is prepended unchanged. Rows retaining fewer than ten residues are discarded. The four procedures differ only in which columns of $A$ receive donor residues and from where.

\begin{algorithm}[h]
\caption{Gapped transplant. Donor residues are written only at structurally corresponding columns.}
\label{alg:gapped}
\begin{algorithmic}[1]
\Require alignment $\mathcal{A}$, donor rows $\mathcal{S}_B$, target length $L_A$
\For{each donor row $s \in \mathcal{S}_B$}
  \State $r \gets$ array of $L_A$ gap characters
  \For{each $(i,j) \in \mathcal{A}$}
    \State $r[i] \gets s[j]$
  \EndFor
  \State emit $r$
\EndFor
\end{algorithmic}
\end{algorithm}

\begin{algorithm}[h]
\caption{Ungapped transplant. The aligned columns are filled as above, and the intervening columns of $A$ receive $B$'s intervening residues.}
\label{alg:ungapped}
\begin{algorithmic}[1]
\Require alignment $\mathcal{A}$ sorted by $i$, donor rows $\mathcal{S}_B$, target length $L_A$
\For{each donor row $s \in \mathcal{S}_B$}
  \State $r \gets$ array of $L_A$ gap characters
  \For{each $(i,j) \in \mathcal{A}$}
    \State $r[i] \gets s[j]$
  \EndFor
  \For{each consecutive pair $(i_1,j_1),(i_2,j_2) \in \mathcal{A}$}
    \State $n_A \gets i_2 - i_1 - 1$; \quad $n_B \gets j_2 - j_1 - 1$
    \If{$n_A > 0$ and $n_B > 0$}
      \State $v \gets (s[j_1{+}1], \dots, s[j_2{-}1])$
      \If{$n_B > n_A$}
        \State discard $n_B - n_A$ entries from the middle of $v$
      \EndIf
      \State write $v$ into $r[i_1{+}1], r[i_1{+}2], \dots$ in order
    \EndIf
  \EndFor
  \State extend outward from the first and last aligned pairs into the flanking columns of $A$,
         taking $B$'s residues in order until either chain is exhausted
  \State emit $r$
\EndFor
\end{algorithmic}
\end{algorithm}

\begin{algorithm}[h]
\caption{Register-shift ablation. The same mapping is displaced $k$ residues along the target chain.}
\label{alg:shift}
\begin{algorithmic}[1]
\Require alignment $\mathcal{A}$, donor rows $\mathcal{S}_B$, target length $L_A$, offset $k$
\State $\mathcal{A}' \gets \{(i+k,\,j) : (i,j) \in \mathcal{A},\; 0 \leq i+k < L_A\}$
\State \Return \Call{GappedTransplant}{$\mathcal{A}'$, $\mathcal{S}_B$, $L_A$}
\end{algorithmic}
\end{algorithm}

\begin{algorithm}[h]
\caption{Register-scramble ablation. The donor's columns land on a random permutation of the same target columns.}
\label{alg:scramble}
\begin{algorithmic}[1]
\Require alignment $\mathcal{A}$, donor rows $\mathcal{S}_B$, target length $L_A$
\State $I \gets (i : (i,j) \in \mathcal{A})$ \Comment{target columns, in order}
\State $J \gets (j : (i,j) \in \mathcal{A})$ \Comment{donor columns, in order}
\State $\pi \gets$ a uniform random permutation of $I$
\State $\mathcal{A}' \gets \{(\pi_m,\,J_m) : m = 1,\dots,|\mathcal{A}|\}$
\State \Return \Call{GappedTransplant}{$\mathcal{A}'$, $\mathcal{S}_B$, $L_A$}
\end{algorithmic}
\end{algorithm}

Both ablations therefore supply the identical donor rows at the identical depth, and occupy the same number of columns as the gapped transplant up to the few positions a shift carries past the end of the chain. They differ from it only in the correspondence between donor and target columns: the shift preserves the donor's residue adjacency while displacing it uniformly, and the scramble destroys adjacency and correspondence together.

\begin{algorithm}[h]
\caption{Corrupted-column ablation. The target's own alignment is retained except at the aligned columns, which receive residues from structurally unaligned positions of the donor.}
\label{alg:corrupt}
\begin{algorithmic}[1]
\Require alignment $\mathcal{A}$, target's own rows $\mathcal{S}_A$, donor rows $\mathcal{S}_B$, donor length $L_B$
\State $I \gets$ sorted target columns of $\mathcal{A}$
\State $F \gets (j : 0 \leq j < L_B,\; j \notin \{j : (i,j) \in \mathcal{A}\})$ \Comment{donor positions outside the alignment}
\For{$n = 1$ to $|\mathcal{S}_A|$}
  \State $r \gets \mathcal{S}_A[n]$ \Comment{the target's own row, unchanged elsewhere}
  \State $t \gets \mathcal{S}_B[((n-1) \bmod |\mathcal{S}_B|) + 1]$
  \For{$m = 1$ to $|I|$}
    \State $r[I_m] \gets t[F_{((m-1) \bmod |F|) + 1}]$
  \EndFor
  \State emit $r$
\EndFor
\end{algorithmic}
\end{algorithm}